\documentclass[12pt,preprint]{aastex}
\usepackage{lscape} 

\def\be{\begin{equation}}
\def\ee{\end{equation}}

\def\m{~$\mu$m}
\def\CI   {[\ion{C}{1}]}
\def\CII  {[\ion{C}{2}]}
\def\HI   {\ion{H}{1}}
\def\HII  {\ion{H}{2}}

\def\NII  {[\ion{N}{2}]}
\def\NII  {[\ion{N}{2}]}
\def\NIII {[\ion{N}{3}]}

\def\OI   {[\ion{O}{1}]}

\def\OIII {[\ion{O}{3}]}

\begin {document}

\title{A Compendium of Far-Infrared Line and Continuum Emission for 227 Galaxies Observed by the Infrared Space Observatory}
\shorttitle{A Compendium of ISO LWS Extragalactic Data}

\author{James~R.~Brauher\altaffilmark{1}, Daniel~A.~Dale\altaffilmark{2}, George~Helou\altaffilmark{1}}
\altaffiltext{1}{\scriptsize California Institute of Technology, MC 314-6, Pasadena, CA 91101}
\altaffiltext{2}{Department of Physics and Astronomy, University of Wyoming, Laramie, WY 82071; ddale@uwyo.edu}

\begin {abstract}
Far-infrared line and continuum fluxes are presented for a sample of 227 galaxies observed with the Long Wavelength Spectrometer on the {\it Infrared Space Observatory}.  The galaxy sample includes normal star-forming systems, starbursts, and active galactic nuclei covering a wide range of colors and morphologies.  The dataset spans some 1300 line fluxes, 600 line upper limits, and 800 continuum fluxes.  Several fine structure emission lines are detected that arise in either photodissociation or \HII\ regions: \OIII~52\m, \NIII~57\m, \OI~63\m, \OIII~88\m, \NII~122\m, \OI~145\m, and \CII~158\m.  Molecular lines such as OH at 53\m, 79\m, 84\m, 119\m, and 163\m, and H$_2$O at 58\m, 66\m, 75\m, 101\m, and 108\m\ are also detected in some galaxies.  In addition to those lines emitted by the target galaxies, serendipitous detections of Milky Way \CII~158\m\ and an unidentified line near 74\m\ in NGC~1068 are also reported.  Finally, continuum fluxes at 52\m, 57\m, 63\m, 88\m, 122\m, 145\m, 158\m, and 170\m\ are derived for a subset of galaxies in which the far-infrared emission is contained within the $\sim$75\arcsec\ ISO LWS beam.  The statistics of this large database of continuum and line fluxes, including trends in line ratios with the far-infrared color and infrared-to-optical ratio, are explored.
\end {abstract}

 
\section {Introduction}
	
Far-infrared wavelengths provide the opportunity to observe dust-enshrouded galaxies without large extinction effects and offer many diagnostics of the physical conditions in the interstellar medium of these galaxies.  The {\it Kuiper Airborne Observatory} provided early data on the far-infrared fine structure lines that arise in photodissociation regions (PDRs) and \HII\ regions in galaxies.  
With the launch of the {\it Infrared Space Observatory} ({\it ISO}; Kessler et al. 1996; Kessler et al. 2003) the far-infrared properties of galaxies were observed with greater sensitivity than ever before.  The Long Wavelength Spectrometer (LWS; Clegg et al. 1996; Gry et al. 2003) on {\it ISO} allowed the large-scale study of far-infrared atomic and molecular lines that supply new insight into the understanding of the interstellar medium of these sources.  Our next opportunities for far-infrared spectroscopic studies of galaxies will come with the {\it Stratospheric Observatory for Infrared Astronomy} and the {\it Herschel Space Observatory}.
	
The LWS data presented in this paper were taken from the {\it ISO} archive.\footnote{http://www.iso.vilspa.esa.es/ida/index.html}  A variety of extragalactic observing programs used the LWS to obtain spectra of the primary diagnostic lines of the interstellar medium in the far-infrared.  These lines include \OIII~52\m, \NIII~57\m, \OI~63\m, \OIII~88\m, \NII~122\m, \OI~145\m, and \CII~158\m.  Among these atomic and ionic fine structure lines, \CII~158\m\ and \OI~63\m\ are the dominant cooling lines for neutral interstellar gas.
Observations of \CII~158\m\ in NGC~6946 (Madden et al. 1993; Contursi et al. 2002) suggest that a significant fraction of the \CII~158\m\ emission might also originate in diffuse ionized gas in some galaxies, while the far-infrared emission lines from ionized species (\OIII~52\m, \NIII~57\m, \OIII~88\m, and \NII~122\m) predominantly originate in \HII\ regions (see also Sauvage, Tuffs, \& Popescu 2005).  Combined with models of PDRs and \HII\ regions (e.g., Tielens \& Hollenbach 1985; Rubin 1985; Wolfire, Tielens \& Hollenbach 1990; Hollenbach, Takahashi, \& Tielens 1991; Spinoglio \& Malkan 1992; Rubin et al. 1994; Kaufman et al. 1999; Abel et al. 2005; Le Petit et al. 2006; Meijerink, Spaans, Israel 2007; Groves et al. 2008), these fine structure transitions can be used to derive gas temperatures, densities, and the intensity of the radiation fields in galaxies.  The LWS was also used to observe a suite of molecular lines in galaxies (Fischer et al. 1999) including hydroxyl (OH; 53\m, 65\m, 79\m, 84\m, 119\m, 163\m), water (H$_2$O; 59\m, 67\m, 75\m, 101\m, 108\m), and the LWS range contains a plethora of high level rotational lines of carbon monoxide (Varberg \& Evenson 1992).  From the detections of multiple transitions of these molecules, the column densities and abundances for OH, H$_2$O, and CO can be determined (e.g., Skinner et al. 1997; Gonz\'alez-Alfonso et al. 2004; Spinoglio et al. 2005; Gonz\'alez-Alfonso et al. 2008).  
	
This contribution reports on LWS observations of seven far-infrared fine structure atomic and ionic lines, far-infrared lines from three molecular species, and the far-infrared continuum of 227 galaxies, in addition to serendipitous detections of Milky Way \CII~158\m\ emission.  The collection of far-infrared line fluxes in this paper comprise the largest sample ever assembled and reduced in a uniform manner.  These line fluxes are used to compare the relationship of the far-infrared fine structure lines, normalized to either another far-infrared line or the far-infrared continuum level, to two indicators of star formation activity: the 60\m/100\m\ ratio and far-infrared-to-$B$ ratio.  The properties of these emission lines are compared to findings from previous LWS emission line studies (Malhotra et al. 1997, 2001; Leech et al. 1999; Fischer et al. 1999; Luhman et al. 1998; Negishi et al. 2001; Luhman et al. 2003).  The LWS continuum fluxes derived in this work are compared to {\it IRAS} 60\m\ and 100\m\ fluxes, ISOPHOT 170\m\ fluxes (Stickel et al. 2000), and infrared spectral energy distribution models for normal star-forming galaxies (Dale \& Helou 2002).  

These data can form an important framework for studies of global extragalactic interstellar media including the derivation of average gas temperatures, densities, abundances, and radiation fields integrated over entire galaxy systems.
The line and continuum fluxes presented here can also supply the data for studies of the individual components (\HII\ regions, spiral arms, disk regions) of large galaxies resolved by the LWS.  Contursi et al. (2002), for example, examine the physical conditions of these different galaxy components in NGC~1313 and NGC~6946 using PDR models (Kaufman et al. 1999), and Johnson et al. (in preparation) explore the relationships between the far- and mid-infrared cooling lines observed respectively by {\it ISO} and the {\it Spitzer Space Telescope}.  LWS studies of individual galaxies have also been carried out for NGC~4038/4039, M~82, NGC~253, Cen~A, NGC~1068, Arp~220, and Mrk~231 (Fischer et al. 1996; Colbert et al. 1999; Unger et al. 2000; Bradford et al. 1999; Gonz\'alez-Alfonso et al. 2004; Spinoglio et al. 2005; Gonz\'alez-Alfonso et al. 2008).
	
Section~\ref{sec:sample} describes the sample of galaxies, while \S~3 describes the observations and data analysis.  In \S~4, the far-infrared continuum data are presented and assessed from comparisons to {\it IRAS} 60 and 100\m\ data, ISOPHOT 170\m\ data, and galaxy infrared spectral energy distribution models.  The far-infrared line data and properties are presented in \S~5.  In \S~6, the statistical trends seen in the line data are described and these trends are related to those found from previous studies.  A summary of the main results is given in \S~7.  The Appendix provides a description of the extended source correction and how it may be applied to the line and continuum fluxes for sources that are extended compared to the $\sim75$\arcsec\ LWS aperture.

\section{The Sample and Data}
\label{sec:sample}

The ISO LWS sample of galaxies selected from the {\it ISO} archive for this paper is presented in Table~1 and lists the galaxy positions, recession velocities, morphologies, optical sizes, and the flux densities of these galaxies in the four {\it IRAS} bands along with the {\it IRAS} 60\m/100\m\ ratios.  The positions, optical sizes, and velocities were taken from the NASA Extragalactic Database (NED) in mid-2004.  
	
The sample includes both normal and Seyfert galaxies that were initially selected by identifying galaxies in the {\it IRAS} Cataloged Galaxies and Quasars Observed in the {\it IRAS} Survey (CGQ; Fullmer \& Lonsdale 1989).  The galaxies identified from the CGQ range in 60\m\ and 100\m\ flux density from 1 to 1300~Jy.  The {\it ISO} Data Archive was queried using this list, from which 198 galaxies were observed in the LWS L01 ``spectral range'' or L02 ``line'' observing modes.  Later, galaxies with {\it IRAS} fluxes less than 1~Jy or those with no cataloged {\it IRAS} flux were added to the sample in order to enlarge the sample.  With these considerations, another 29 galaxies were identified within the {\it ISO} Archive.  Photometric mode L02 observations in which the grating remained in a fixed position are excluded from this sample.  The large, nearby galaxies M~31 and the Small and Large Magellanic Clouds are excluded from this sample because the size of these three galaxies is over 100 times larger than the LWS beam.  

Among these 227 galaxies there are two distinct subsets, distinguished by the far-infrared size of the galaxy.  The 181 galaxies in the first subset are unresolved in the far-infrared with respect to the $\sim$75\arcsec\ LWS beam.  This unresolved subset of galaxies is an extension of the combined sets of smaller samples observed with the LWS (Malhotra et al. 2001; Pierini et al. 1998; Luhman et al. 2003; Negishi et al. 2001) with additional sources added from the {\it ISO} Data Archive.  The second subset consists of 46 galaxies resolved by the LWS beam in the far-infrared.  The data from this resolved subset of galaxies can be used to complement past studies (Stacey et al. 1991; Madden et al. 1993, 1997) of large galaxies with data taken from the {\it Kuiper Airborne Observatory} and {\it ISO}.  The resolved galaxies are denoted as such in Table~1.  The {\it IRAS} flux densities presented in Table~1 are selected from either Rice et al. (1988) or Dale et al. (2000) for the large, nearby galaxies in the resolved subset.  For the unresolved subset of galaxies, the {\it IRAS} fluxes in Table~1 are taken from SCANPI co-additions of the {\it IRAS} survey scans.   

The galaxies in this sample are distributed across the entire sky.  Figure~\ref{fig:aitoff} displays the galaxy distribution in Galactic coordinates.  The clump of galaxies at ($l,b$)$\sim$(280\degr,74\degr) is the Virgo Cluster.  Approximately 40 galaxies lie within the Zone of Avoidance, where $|$b$| < 20$\degr.  Although there may be some serendipitous Galactic line and continuum emission in all directions of the sky, this serendipitous contamination is more likely toward galaxies within the Zone of Avoidance in either the galaxy spectra or the spectra of an off-source position taken during these observations.  This contamination of the observed line measurements by Milky Way emission is a concern, and a discussion of detected Galactic emission lines is found in Section~5.

\begin{figure}
 \plotone{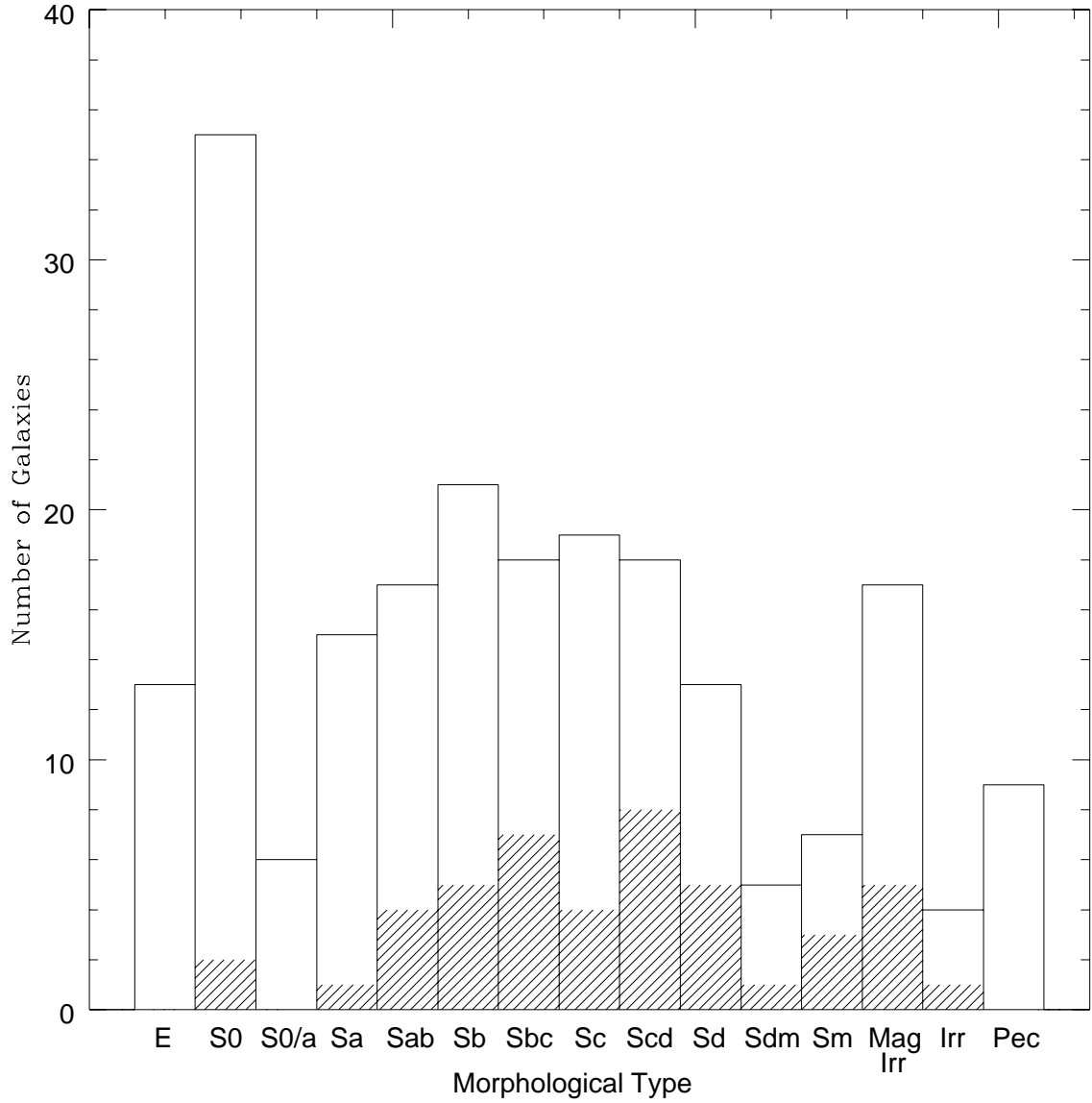}
 \caption{The distributions of the optical morphologies of the resolved and unresolved subsets of galaxies.  The resolved subset is cross-shaded.}
\label{fig:morphology}
\end{figure}

The {\it Palomar Observatory Sky Survey} plates and other observations found within NED were used to reexamine the optical morphology for each galaxy (Harold Corwin, private communication).  In Figure~\ref{fig:morphology}, the distribution of the optical morphological types for the two subsets is shown.  Both the resolved and unresolved subsets span the range of early- to late-type galaxies.  The unresolved subset contains a relatively large number of S0 galaxies while the resolved subset contains no elliptical, S0/a, or peculiar galaxies.

\begin{figure}
 \plotone{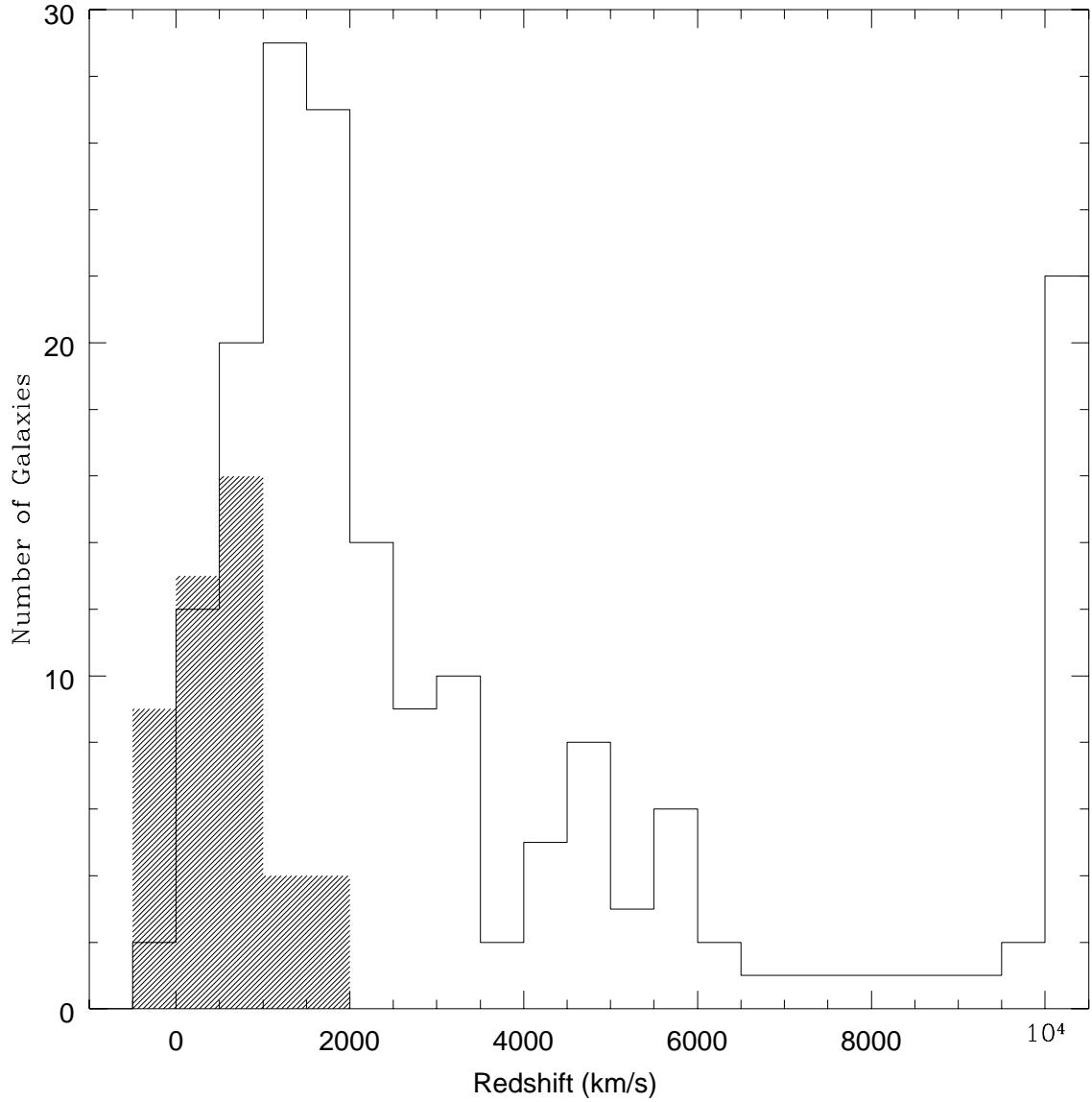}
 \caption{The distributions of redshifts for the resolved and unresolved subsets.  The resolved galaxy redshifts are cross-shaded.}
\label{fig:cz}
\end{figure}

Most of the galaxies in both the unresolved and resolved subsets are relatively nearby, and Figure~\ref{fig:cz} shows the redshift distribution for both subsets.  All galaxies in the resolved subset have an absolute redshift less than 2000~km~s$^{-1}$.  With the exception of a large bin (22/181) of galaxies with redshifts greater than 10000~km~s$^{-1}$, most galaxies in the unresolved subset have redshifts less than 6000~km~s$^{-1}$.

\begin{figure}
 \plotone{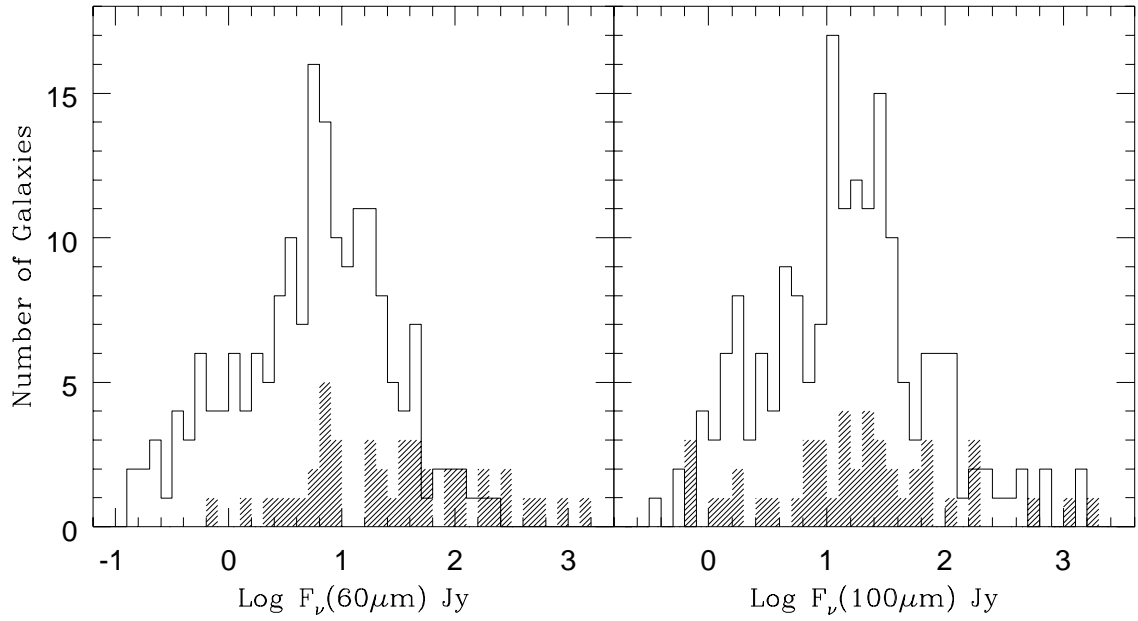}
 \caption{The distributions of {\it IRAS} 60\m\ and 100\m\ flux densities.  The resolved galaxy subset is cross-shaded.}
\label{fig:iras_fluxes}
\end{figure}

Figure~\ref{fig:iras_fluxes} shows the distribution of measured flux densities at 60\m\ and 100\m\ for the resolved and unresolved galaxies in this contribution, with {\it IRAS} detections taken from Table~1.  At 60\m, the resolved galaxies span a more elevated flux density range ($\sim 1-1300$~Jy) than the unresolved galaxies ($\sim0.2-150$~Jy), with the median flux density for unresolved galaxies smaller by a factor of roughly 4.5.  At 100\m, the unresolved and resolved galaxy subsets have similar distributions over flux density, covering the range $\sim$0.5 to 1000~Jy with the median flux density of unresolved galaxies smaller by only a factor of 1.6.  

\begin{figure}
 \plotone{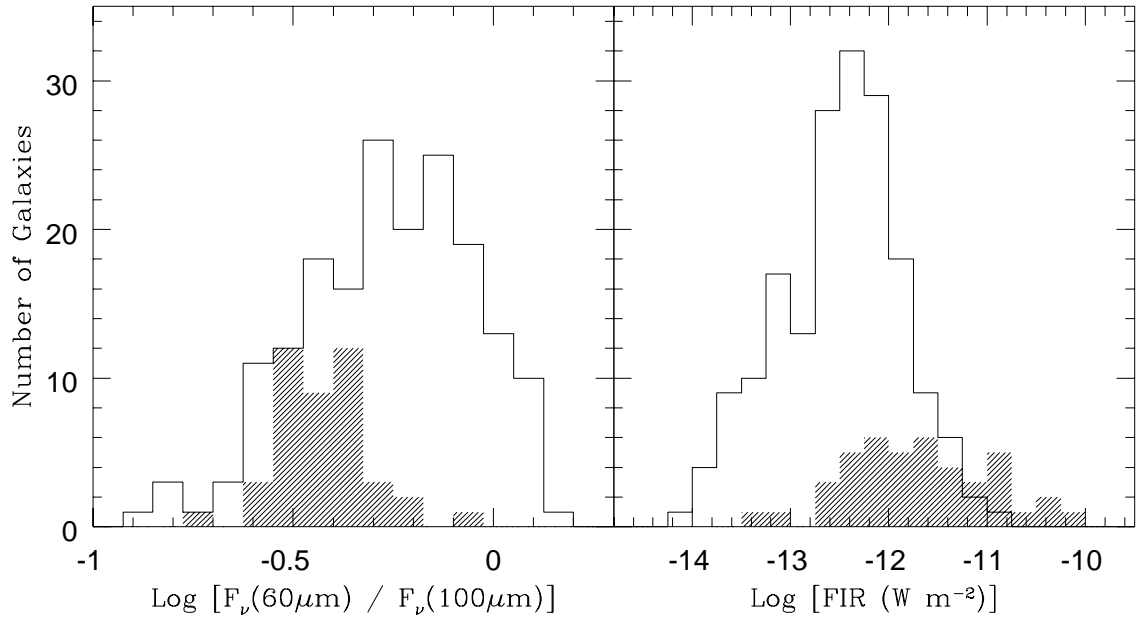}
 \caption{The distributions of {\it IRAS} 60\m/100\m\ ratio and {\it FIR}.  The resolved galaxy subset is cross-shaded.}
\label{fig:iras_colors}
\end{figure}

\begin{figure}
 \plotone{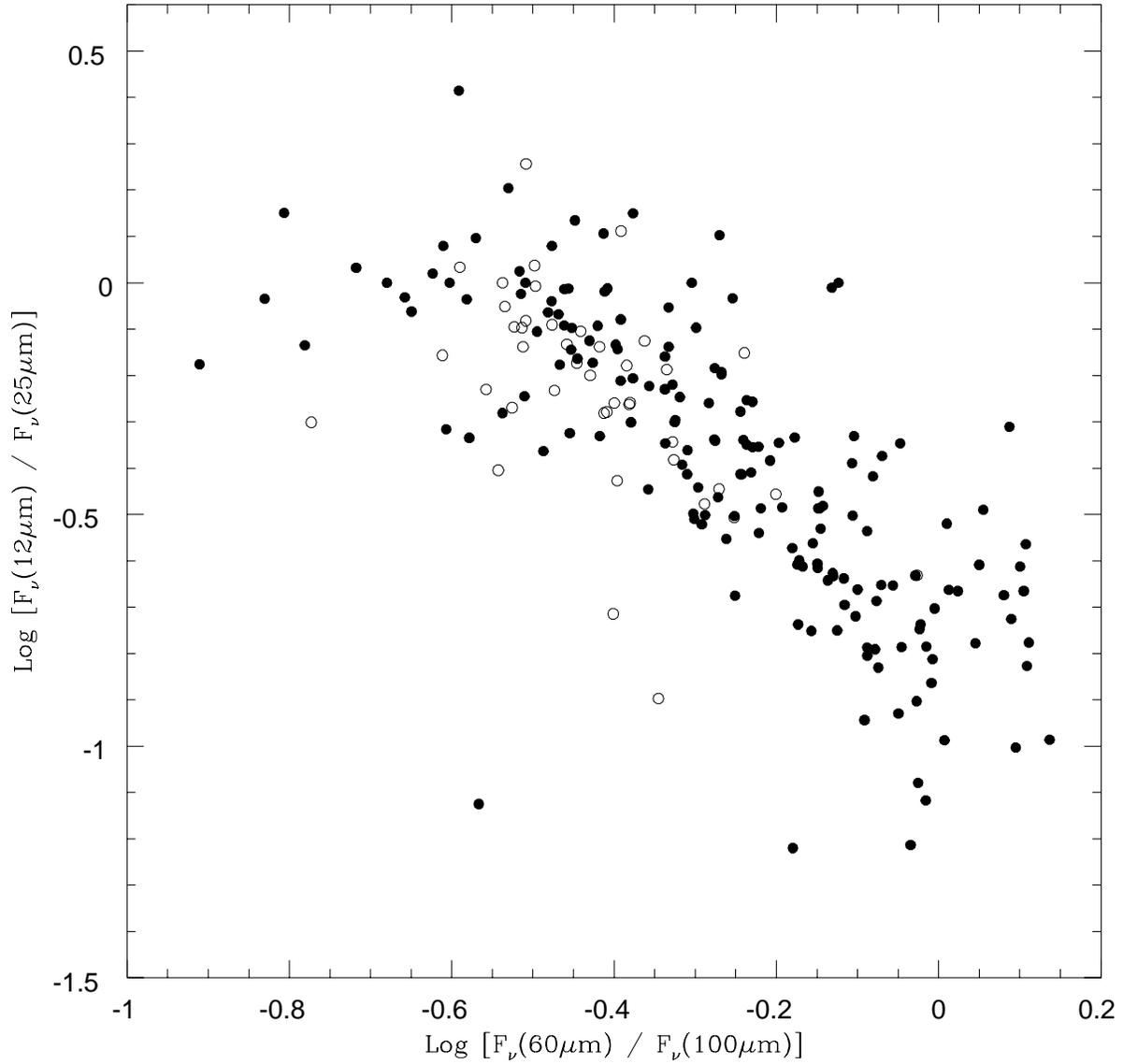}
 \caption{The {\it IRAS} color-color diagram for the galaxies in this sample.  Quiescent galaxies are located towards the upper left and actively star-forming galaxies towards the lower right.  Filled circles are galaxies in the unresolved subset.  Open circles are galaxies in the resolved subset.}
\label{fig:iras_color_color}
\end{figure}

The distributions of the 60\m/100\m\ ratio and far-infrared flux are displayed in Figure~\ref{fig:iras_colors}.  The 60\m/100\m\ ratio is an indicator of the typical heating intensity of dust in galaxies and may also suggest the relative star formation activity level of a galaxy.  Lower 60\m/100\m\ ratios typically correspond to quiescent galaxies, whereas higher 60\m/100\m\ ratios indicate either a higher rate of star formation or perhaps the presence of an AGN (Helou 1986).  The far-infrared flux is defined as {\it FIR}~$= 1.26 \cdot 10^{-14} [2.58 f_\nu(60\micron) + f_\nu(100\micron)]$~W~m$^{-2}$, where $f_\nu(60$\m) and $f_\nu(100$\m) are the 60\m\ and 100\m\ {\it IRAS} flux densities in Jy (Helou et al. 1988).  The unresolved and resolved galaxy subsets peak near 60\m/100\m\ ratios of 0.5 and 0.4, respectively, and the resolved subset does not contain many warm galaxies with 60\m/100\m\ ratios greater than 0.7.  It is not surprising that resolved galaxies are on average closer and more quiescent than unresolved galaxies.  The larger subset of unresolved ($\sim$distant) systems should include more galaxies exhibiting extreme luminosities and activity levels.  The distribution of {\it FIR} values for the two subsets spans five orders of magnitude with a peak between 10$^{-12}$ and 10$^{-13}$~W~m$^{-2}$.  The resolved galaxies reach {\it FIR} values as large as 10$^{-10}$~W~m$^{-2}$, an order of magnitude larger than the brightest unresolved galaxies.  In Figure~\ref{fig:iras_color_color}, the {\it IRAS} 12\m/25\m\ ratio is plotted against the 60\m/100\m\ ratio for the resolved and unresolved subsets.  The sequence of infrared colors in Figure~\ref{fig:iras_color_color} is associated with a sequence of star formation activity in galaxies (Helou 1986) and dust-heating intensity (Boulanger et al. 1988), with the upper left populated by quiescent galaxies and the lower right by warmer, more actively star-forming galaxies.

\section{Observations and Data Analysis}
Observations were made using the LWS in grating mode (L01, L02, 43--197\m, $\lambda/\Delta \lambda \sim 200$).  The LWS consists of ten detectors with spectral overlap for adjacent detectors.   In the grating mode of the LWS, the spectral resolution is about 0.29\m\ for the 10\m-wide short-wavelength detectors (SW1-SW5) and 0.60\m\ in the 20\m-wide long-wavelength detectors (LW1--LW5).  The L01 Astronomical Observation Template (AOT) is a range scan of the grating that results in 10 spectra covering a significant range of the LWS.  The L02 AOT produces spectra for up to ten wavelengths, specified by the observer.  In this mode, data are recorded for all ten detectors while the specified wavelengths are being scanned, producing spectra with significant gaps across the range of the LWS.   

All guaranteed and open time observations for 227 galaxies were extracted from the {\it ISO} Data Archive and processed through the LWS Pipeline Version 7.0 or 8.7.  Slight improvements in the photometric model are made beyond Pipeline Version~7.0, but these changes have minor effects on the calibration of the L01 and L02 grating mode AOTs.  These changes yield improvements in the flux accuracies by a few percent but do not significantly alter the line and continuum fluxes that are derived from Pipeline~7.0.     

Further data manipulation is then carried out using the LWS Interactive Analysis (LIA; Hutchinson et al. 2001) and the {\it ISO} Spectral Analysis Package (ISAP; Sturm et al. 1998).  The continuum fluxes in the LWS spectra are significantly affected by the uncertainties in the dark current, which can be of the same order as the source continuum.  Many of the galaxies in this sample are in this faint flux regime ($f_\nu(60\micron) < 50$~Jy in the 75\arcsec\ LWS beam).  As the dark currents are only additive in nature across the whole band, they do not affect the line flux estimates.  The dark currents are re-estimated and removed one at a time by hand through visual inspection using the LIA.  The data are then corrected detector by detector for any evident instrumental responsivity variations and flux calibrated to the LWS calibration source Uranus, applied using LIA.  Glitches due to cosmic rays are removed by hand from the data using ISAP by plotting spectral scans as a function of time and identifying bad data points through the characteristic appearance of falling glitch trails.  Depending on the quality of the observation of a galaxy, between 15\% and 20\% of the data are typically discarded.  Spectral scans are co-added and averaged together using a 3$\sigma$ clip in spectral bins of about 0.05\m.  For extended sources or for sources that are off-center with respect to the LWS aperture, a sinusoidal fringe associated with internal reflection and interference within the LWS instrument may arise (Gry et al. 2003; Swinyard et al. 1998).  The fringes are usually less than 5\% of the continuum and do not severely affect the line and continuum measurements.  For full-grating L01 observations, these fringes can be removed using a defringing algorithm available within ISAP.  The LWS data also suffer from transients.  When the grating is scanned between the forward and reverse directions, a small ($<$5\%) detector memory effect (Gry et al. 2003) may be visible between the two scan directions.  This memory effect is due to different response times for the detectors depending on whether the signal increases or decreases with time and is most visible in the SW1, SW2, and LW2 detectors during L01 observations.  No correction is applied for these memory effects.  When these memory effects are present in the data, each scan direction is averaged separately, and the line and continuum fluxes for each scan direction are measured before estimating the final fluxes and uncertainties (see Sections~4 and 5 for further details).  An additional source of uncertainty occurs for extended sources where the variation in the LWS beam from detector to detector might cause a mismatch between adjacent detectors by up to 30\% depending on the extent and structure of the galaxy.  With the application of an extended source correction, this mismatch can be partially corrected.  The data presented in this paper are based on the point source calibration of the pipeline and no correction for extended sources has been applied due to the uncertainty in this correction.  See the Appendix for the definition and discussion of the extended source correction.

\begin{figure}
 \plotone{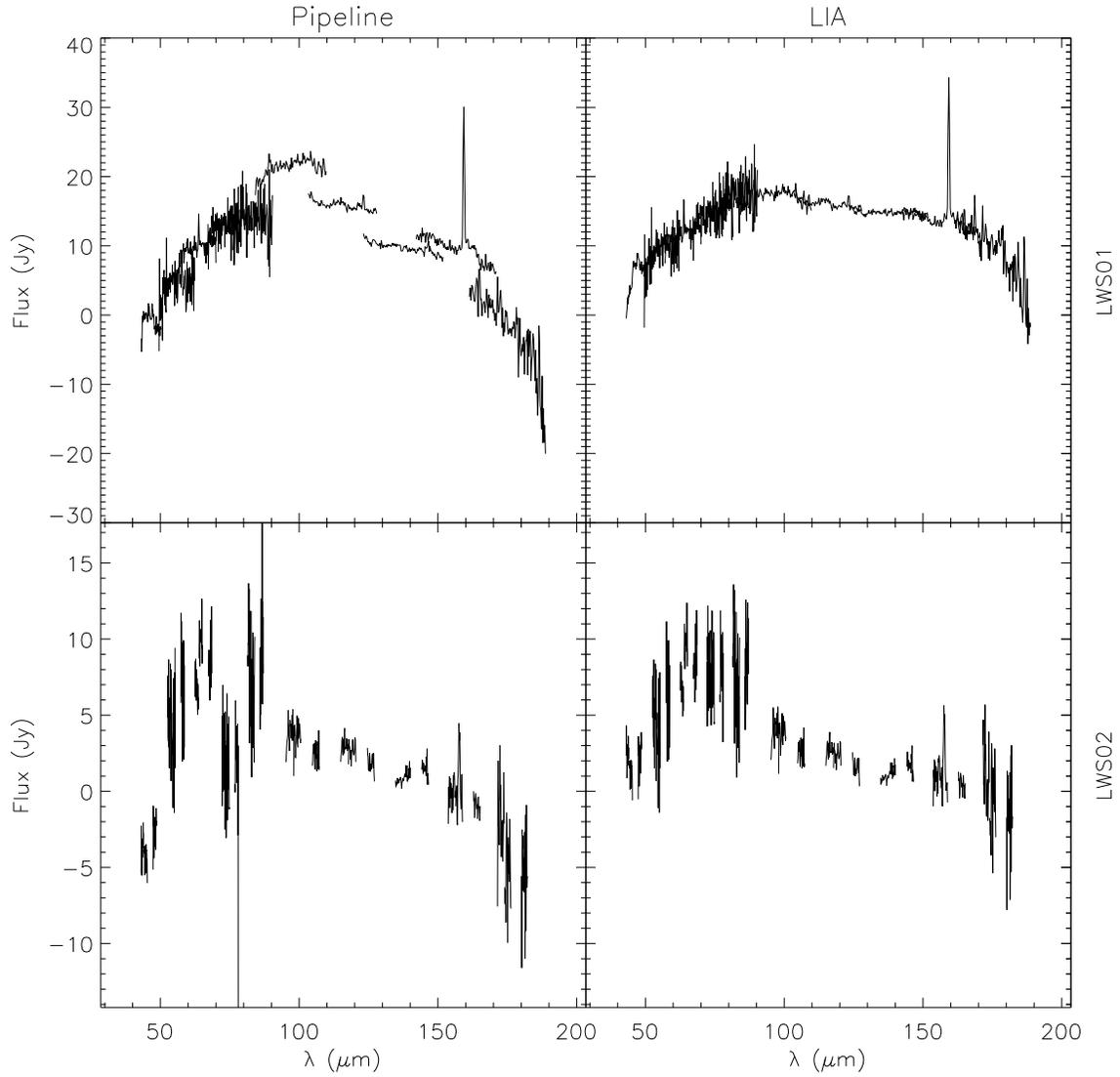}
 \caption{Two example LWS spectra representing the pipeline product L01 and L02 AOTs displayed before and after using the LIA and ISAP reduction.  Improvements from using LIA and ISAP include not only the removal of fringes and glitch removal but also the reduction of negative continuum fluxes and misaligned adjacent detectors.}
\label{fig:processing}
\end{figure}

Through the use of LIA and ISAP, the improvement in the overall quality of the data from the original pipeline Auto-Analysis Result product is substantial.  By re-estimating the dark currents, the appearance of negative fluxes in most of these observations is removed.  Through the re-estimation of the dark currents and gain corrections and careful glitch removal, the match between overlapping detectors is improved, thus producing more continuous spectra, shown in Figure~\ref{fig:processing}.  Any remaining spectral mismatch between adjacent detectors may be the result of residual errors in the dark current subtraction or beam uncertainties from detector to detector.  Using LIA and ISAP, the line and continuum calibration uncertainties decrease from 20\%--30\% to 10\%--20\%, on average, for faint sources ($f_\nu(60\micron) < 50$~Jy) as illustrated in Figure~\ref{fig:processing}.  

\section{The Continuum Data}

The ``monochromatic'' continuum fluxes are derived from the LWS spectra by fitting a 2--5\m\ linear baseline through the spectra surrounding the wavelengths 52\m, 57\m, 63\m, 88\m, 122\m, 145\m, 158\m, and 170\m.  Continuum fluxes are measured only for the well-calibrated LWS spectra when spectra are available at these wavelengths.  Continuum fluxes could not be derived in some L02 observations that had no spectra at these wavelengths.  If the observations were affected by changes in the responsivity and dark current caused by warm-ups in the long wavelength detectors, no continuum flux is derived.  These continuum fluxes and associated uncertainties are listed in Table~2.  Because of the uncertainty of off-axis continuum contributions in extended sources, only fluxes for galaxies unresolved by the LWS beam are listed.
	
Since the L02 observations do not have much overlap between detectors, it is difficult to properly estimate the dark current, and thus the continuum, for faint sources.  In order to test the consistency of the continuum of the L02 observations, the continuum data are compared with observations of the 158\m\ and 63\m\ lines for galaxies that had a pointing at the same position in both the L01 and L02 AOTs.  The continuum flux is taken from the linear baseline that is used for the fit to the lines at 158\m\ and 63\m\ for these observations.  The continuum in the L02 observations accurately reproduces the continuum measured in the L01 AOT to within 5\% for continuum fluxes down to 10~Jy.  The continuum correlation between the two AOTs holds for fluxes below 10~Jy, but the dispersion in this relation increases by a factor of two.  Thus, the L02 continuum appears to be consistent with the L01 continuum although some uncertainties exist from the dark or gain calibration due to the lack of overlapping spectra from adjacent detectors.  
	
The continuum uncertainties quoted in Table~2, typically 15--20\%, are a combination of the measurement and calibration uncertainties.  In most observations, the calibration uncertainties are the dominant source of uncertainty in the continuum fluxes, but in the low flux limit ($<$10~Jy), the measurement uncertainties become dominated by uncertainties in the dark current.  The detector dark currents are of the same order as the continuum in this flux regime.  The effect of these uncertainties in the dark current for continuum levels below 10~Jy is described below in Sections~\ref{sec:vs_iras} and \ref{sec:vs_isophot}.  Although the comparisons to {\it IRAS} and ISOPHOT data show an excellent overall agreement with the LWS continuum fluxes within the LWS 20\% uncertainties down to fluxes below 10~Jy, the dispersion in these relationships increases by as much as a factor of two below 10~Jy.  Although there may be large uncertainties, the measured continuum fluxes below 10~Jy are included in Table~2 because there are no biases in the agreement of the LWS fluxes with {\it IRAS} and ISOPHOT in this low-flux limit.

\subsection{Comparison with {\it IRAS} 60 and 100\m\ Data}
\label{sec:vs_iras}
An extensive comparison of the LWS continuum fluxes and {\it IRAS} catalog fluxes was carried out for galaxies unresolved by the LWS.  The following criteria were applied to this study:  1) galaxies must have an {\it IRAS} Point-Source Catalog detection, 2) the far-infrared emission must be concentrated within the 75\arcsec\ LWS beam, and 3) LWS data must exist in the vicinity of 60\m\ and 100\m\ wavelengths to be used for a continuum estimation.  Using these three criteria, 41 galaxies were observed in the L01 AOT and 104 galaxies in the L02 AOT.     
	
For L01 observations, the 60\m\ and 100\m\ continua are estimated by performing synthetic photometry with an algorithm provided in ISAP.  This photometry is performed on the spectra by integrating across the {\it IRAS} passbands.  The 60\m\ and 100\m\ flux densities are then derived using the {\it IRAS} assumption that the source spectral energy distribution is of the form $\lambda f_\lambda =$ constant.  For the comparison at 60\m\ a correction must be introduced because the LWS does not cover the entire {\it IRAS} 60\m\ filter (43--197\m\ vs. 27--87\m).  The galaxies used for this comparison span a wide range of 60\m/100\m\ ratios (0.2--1.4), so the amount of the integrated flux in the {\it IRAS} 60\m\ filter missed by the LWS varies depending on the shape of the infrared spectral energy distribution of the galaxy.  Dale \& Helou (2002) present a sequence of galaxy spectral energy distributions sorted across a range of 60\m/100\m\ ratios.  Using these models, the amount of the total integrated flux in the {\it IRAS} 60\m\ filter missed by the LWS between 27\m\ and 43\m\ varies from 3\% to 7\% for these 60\m/100\m\ ratios.  This correction based on the 60\m/100\m\ ratio of the galaxy is applied to the LWS 60\m\ fluxes derived from L01 observations.       

For L02 observations, synthetic photometry cannot be used because of the gaps between the multiple short scan spectra.  Instead, a monochromatic flux is estimated by fitting a linear baseline to a 2--5\m\ slice of spectra surrounding the wavelengths 60\m\ and 100\m.  
The L02 60\m\ and 100\m\ monochromatic continuum fluxes must also have a secondary correction applied because these monochromatic fluxes cannot be directly compared to the {\it IRAS} fluxes.  Since the fluxes derived from the L02 observations are taken by fitting a linear baseline through the {\it IRAS} filter central wavelengths, the assumption would be that the flux at these wavelengths is equal to the flux over the entire {\it IRAS} passband for galaxies of these 60\m/100\m\ colors.  A calculation of the difference between the monochromatic fluxes, estimated by fitting a linear baseline to the spectra at 60\m\ and 100\m\ and the synthetic photometry fluxes from integrating over the {\it IRAS} filters, is done using a set of pointed observations in which the same sky position on a galaxy was observed in both the L01 and L02 AOTs.  The monochromatic fluxes overestimate the integrated spectral photometry on average by 11\% at 60\m\ and 9\% at 100\m.  The monochromatic L02 fluxes are corrected for these overestimations from using linear fits to the spectra.                     
	
\begin{figure}
 \plotone{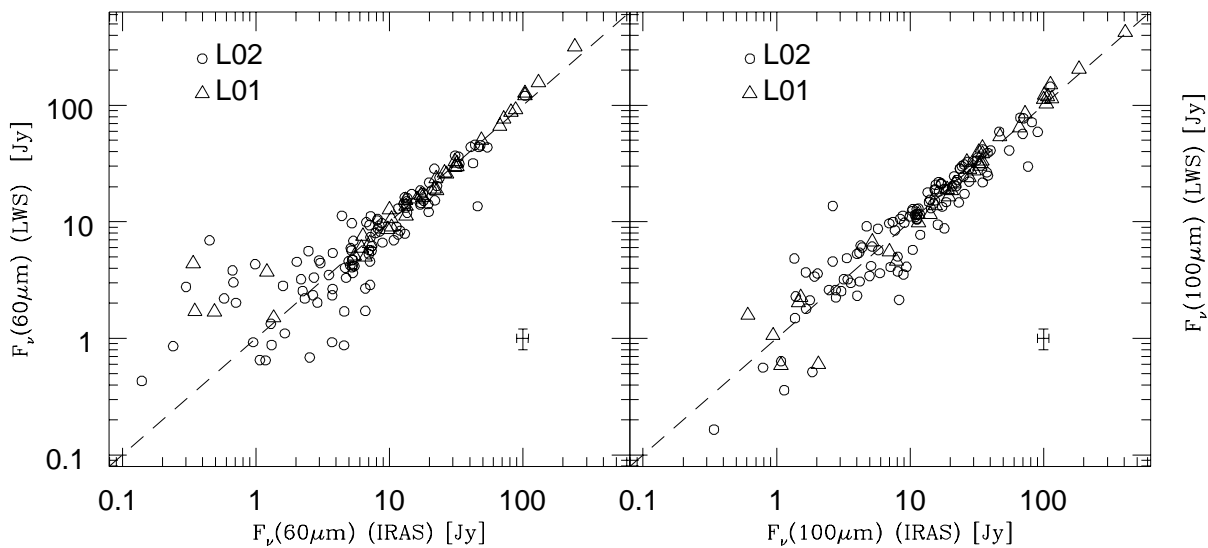}
 \caption{Comparison of LWS continuum fluxes to the {\it IRAS} fluxes at 60\m\ and 100\m.  Triangles (circles) are continuum fluxes from L01 (L02) observations.  A typical error bar is plotted in the lower right of each plot.  The LWS error bar is calculated from the combination of the systematic and measured flux uncertainties and represents an average 20\% uncertainty.  The {\it IRAS} flux error bar is taken from the {\it IRAS} Point Source Catalog and represents an average 10\% uncertainty.  The dashed line is the one-to-one correlation.}
\label{fig:LWS_IRAS}
\end{figure}

After these two corrections are applied, the background is estimated using IRSKY and then removed from the LWS data.  Figure~\ref{fig:LWS_IRAS} shows a comparison of the LWS and {\it IRAS} data.  The LWS error bars are a combination of the uncertainties associated with the baseline fit to the line and the pipeline calibration uncertainties.  The {\it IRAS} error bars are taken directly from the {\it IRAS} Point Source Catalog as given by NED.  The LWS fluxes are, on average, 1\% lower and 2\% higher than {\it IRAS} at 60\m\ and 100\m, respectively, for {\it IRAS} fluxes above 10~Jy, a remarkable agreement.  Below 10~Jy, the dispersion increases from 20\% to 50\%, not surprising since the dark current is comparable to 10~Jy for these sources.  Both the L01 and L02 continuum fluxes show similar offsets and dispersions in the LWS--IRAS comparison.
  
\subsection{Comparison with ISOPHOT 170\m\ Continuum Data}
\label{sec:vs_isophot}
For galaxies where the infrared continuum peaks near 60\m, the continuum level at 170\m\ is a factor of 2 or more lower compared to the peak.  For galaxies with 60\m\ flux densities less than 10~Jy, the detection limit of the LWS may be reached, and the reliability of the 170\m\ continuum is once again subject to uncertainties in the dark current.  Using published fluxes from ISOPHOT pointed observations at 180\m\ (Klaas et al. 2001) and 170\m\ Serendipity Survey (Stickel et al. 2000), 25 galaxies constrained to the LWS beam are compared to 170\m\ fluxes measured by the LWS.

Both ISOPHOT studies referenced above utilized the broad C\_160 filter.  Therefore, two corrections must be applied to the LWS monochromatic fluxes.  These monochromatic fluxes are compared to 170\m\ fluxes derived by integrating over the ISOPHOT C\_160 filter using the synthetic photometry algorithm in ISAP.  On average, the monochromatic LWS fluxes are 11\% higher than their synthetic counterparts, and this correction is applied to the LWS monochromatic fluxes.  The second correction applied adjusts for the difference in wavelength spanned by the LWS and ISOPHOT C\_160 filter (43--197\m\ vs. 100--240\m).  The flux missed by the LWS in the ISOPHOT filter depends upon the spectral shape of the galaxy, and this flux is estimated by assuming the Dale \& Helou (2002) spectral shape of a galaxy for a given 60\m/100\m\ ratio.  Typically, this correction is approximately 3\% for the range of 60\m/100\m\ ratios of these galaxies.

\begin{figure}
 \plotone{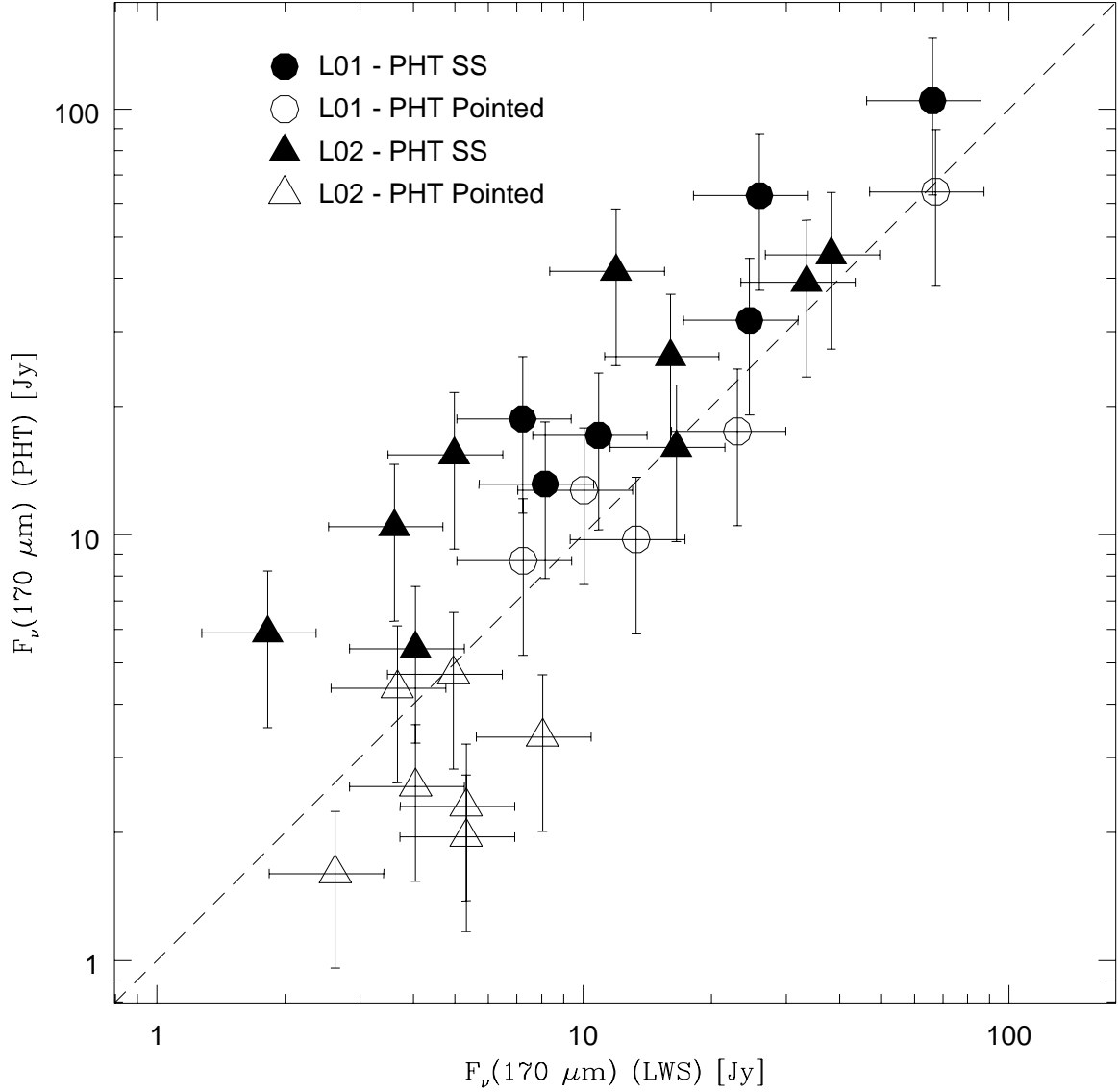}
 \caption{Comparison of LWS continuum fluxes to ISOPHOT Serendipity Survey fluxes (Stickel et al. 2000) at 170\m\ for galaxies unresolved by the LWS.  Filled circles are measurements from L01 observations and open triangles are from L02 observations.  The ISOPHOT error bars represent 40\% calibration uncertainties quoted by Stickel et al. (2000) and the LWS error bars represent 30\% calibration uncertainties at 170\m.  The dashed line is the one-to-one correlation.}
\label{fig:LWS_ISOPHOT}
\end{figure}

The total correction applied to LWS flux densities is approximately 14\%, and Figure~\ref{fig:LWS_ISOPHOT} is the resulting plot of this comparison.  The ISOPHOT and LWS flux densities track each other well between 1 and 100~Jy, particularly if just ISOPHOT Pointed data are considered, although the dispersion increases significantly below 10~Jy.  This effect is due to the large uncertainties in the dark current at this wavelength and flux regime.  The ISOPHOT Serendipity Survey (SS) flux densities are systematically a bit high compared to the LWS data.  Serendipitous Survey sources were observed as they streamed by the instrument field-of-view; the fluxes reconstructed from the glancing scans may have been slightly overestimated.  Considering the differences in beam size and calibration between the LWS and ISOPHOT instruments, the 30\% overall agreement is consistent with the uncertainties of the two instruments at this wavelength.

\subsection{Comparison with Models of Galaxy Infrared Spectral Energy Distributions}
\begin{figure}
 \plotone{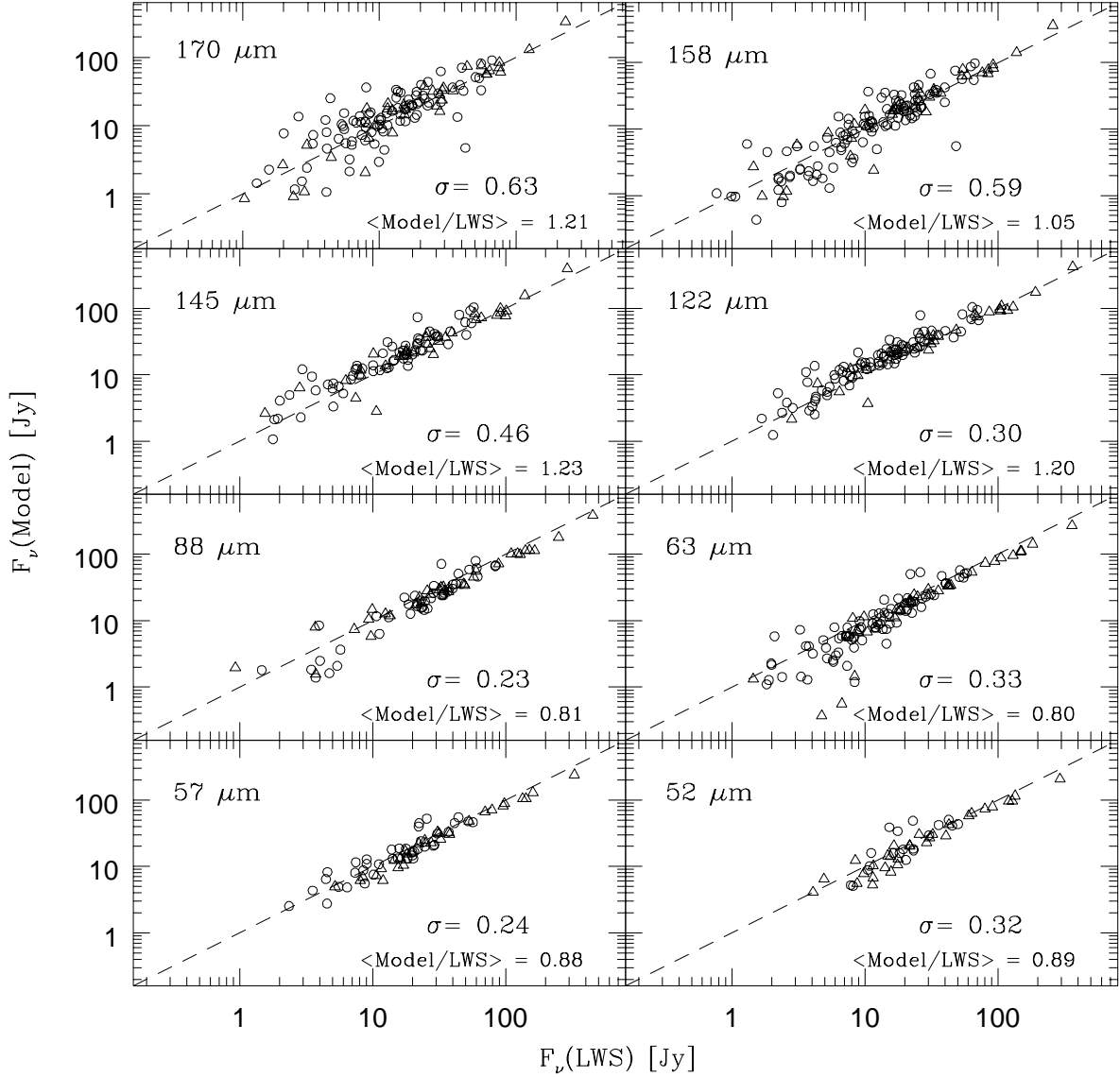}
 \caption{Comparison of the LWS continuum fluxes of galaxies unresolved by the LWS beam and the Dale \& Helou (2002) model prediction for eight far-infrared wavelengths.  Triangles (circles) represent continuum fluxes taken from a fully sampled L01 (fitted L02 line) spectrum.  The dashed line is the one-to-one correlation.}
\label{fig:lws_vs_model}
\end{figure}

The data for the subset of galaxies smaller than the LWS aperture discussed in the previous section are compared to a semi-empirical model for the infrared spectral energy distributions of normal star-forming galaxies between 3 and 1100\m\ (Dale et al. 2001; Dale \& Helou 2002).  The comparison of the LWS and model flux densities provides a consistency check of the LWS continuum flux densities, especially at the longer wavelengths where few continuum measurements from other observatories exist.  The model is based on the combination of emission curves for large and very small grains and aromatic feature carriers for varying interstellar radiation fields, and are combined assuming a variable power-law distribution of dust mass over heating intensity.  The model is constrained by {\it IRAS} and {\it ISO} observations of a sample of 60 normal, star-forming galaxies (Dale et al. 2000).  According to this model, a sequence of global star formation activity level is formed as galaxies are sorted according to their 60\m/100\m\ ratio; the {\it IRAS} 60\m/100\m\ ratios are used in conjunction with the models to predict the LWS continuum levels at various wavelengths.  The observed LWS continuum agrees with the predicted model flux densities (anchored by 60 and 100\m\ {\it IRAS} photometry) for these galaxies to within 25\% at 52\m, 57\m, 63\m, 88\m, 122\m, 145\m, 158\m, and 170\m\ and is presented in Figure~\ref{fig:lws_vs_model}.  The spectral energy distribution models and the comparison with the LWS continuum are explored further in Dale \& Helou (2002).         

\section{The Line Data}

\begin{figure}
 \plotone{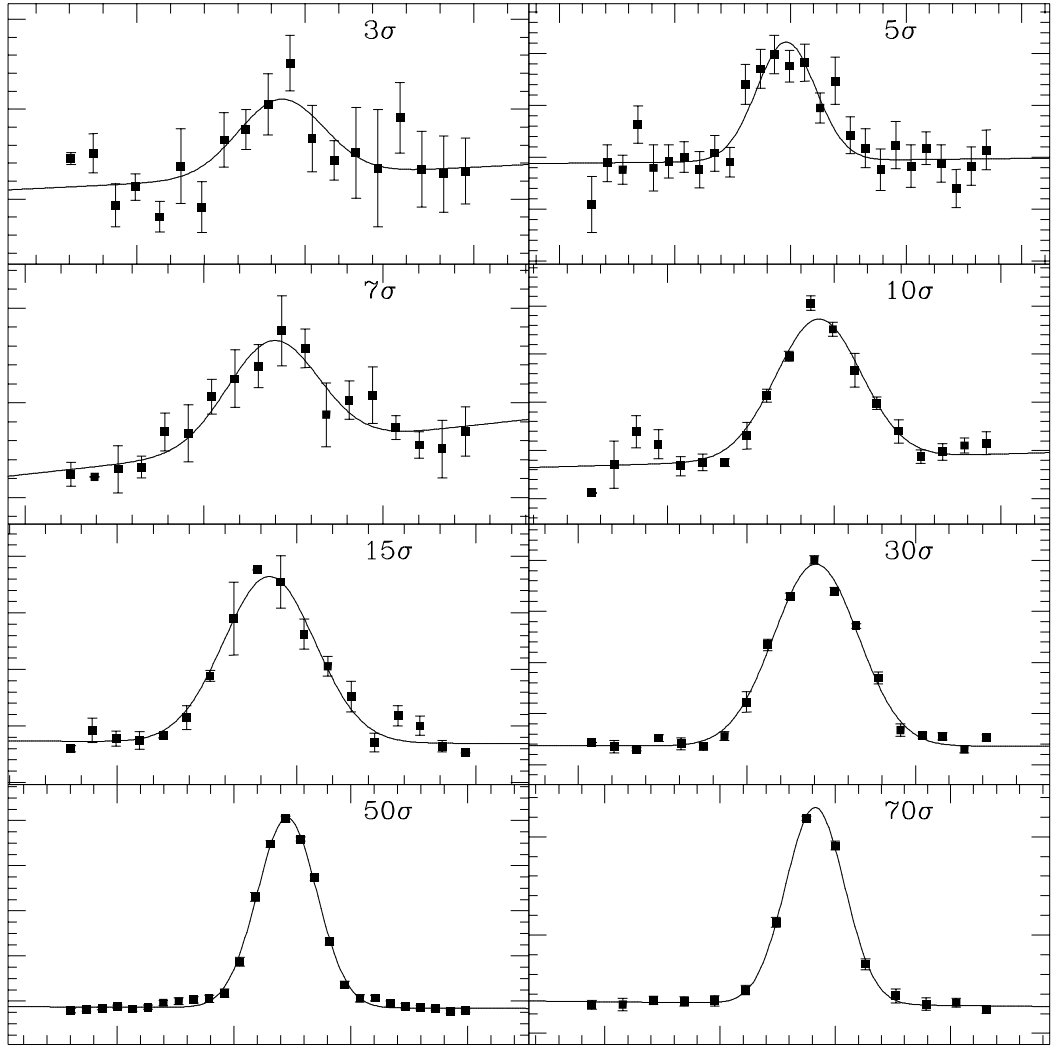}
 \caption{Examples of the \CII~158\m\ line found in this sample.  This figure is a representative selection of varying signal-to-noise detections and shows how well a Gaussian with the effective instrumental width fits the line data.  Error bars are taken from the uncertainty in the averaging of the spectral scans in each bin.}
\label{fig:CII}
\end{figure}

All spectral lines in this paper are unresolved ($\Delta v\sim1500$~km~s$^{-1}$).  Thus, the line fluxes are calculated assuming the line profile to be dominated by a Gaussian instrumental profile (FWHM=0.29\m\ for $\lambda <$ 93\m, 0.60 for $\lambda >$ 80\m; for wavelengths between 80 and 93\m, the spectral resolution depends on which of the two overlapping detectors (SW5, LW1) the line was measured).  A Gaussian has been shown to fit the LWS instrumental profiles to within 2\% (Gry et al. 2003).  Example line scans from this sample with various signal-to-noise ratios are displayed with a Gaussian fit in Figure~\ref{fig:CII}.  A detected line in this sample is defined as one that has a peak flux at the 3$\sigma$ or higher confidence level.  The statistical uncertainty associated with each line is $\Delta \lambda f_\lambda({\rm r.m.s.})$, the spectral resolution times the root mean square variations in the flux density of the local continuum.  The dominant uncertainty for most observations is the systematic flux uncertainty that is taken from the pipeline processing.  This uncertainty is a combination of the dark current, illuminator, and Uranus model calibration uncertainties.  The total uncertainty is calculated by adding the statistical and systematic uncertainties in quadrature.  The total uncertainty is typically between 10\% and 20\% of the line flux measurement, depending on the quality of the observation.  In the case of non-detections, 3$\sigma$ upper limits are calculated by multiplying the local statistical uncertainty describe above by 3.\footnote{Upper limits are also available via L01 range scans, though they are not presented here.}  The data for galaxies from previous studies are included as a subset of the larger sample presented here, but are independently reduced in the manner described in \S~3.  In general, the line fluxes presented here agree with literature values to within 30\%.  The lines, rest wavelengths, and transitions for this sample are listed in Table~3.

\subsection{Far-Infrared Fine Structure Lines}
A number of studies have produced models that predict the strength of far-infrared fine structure lines such as \CII~158\m, \OI~145\m, and \OI~63\m\ as a function of the density and radiation intensity in PDRs (e.g., Tielens \& Hollenbach 1985; Wolfire, Tielens \& Hollenbach 1990; Hollenbach, Takahashi, \& Tielens 1991; Spinoglio \& Malkan 1992; Kaufman et al. 1999; Abel et al. 2005; Le Petit et al. 2006; Dopita et al. 2006; Dopita et al. 2006; Meijerink, Spaans, Israel 2007; Groves et al. 2008).  The \CII~158\m\ and \OI~63\m\ lines act as the primary coolants to the dense ($n \sim 10-10^5$~cm$^{-3}$ or more), warm ($T \sim 100-1000$~K), neutral media.  Other far-infrared fine structure lines probed by {\it ISO}, such as \NII~122\m, \OIII~52\m\ and 88\m, and \NIII~57\m, are important to understanding \HII\ regions.  From these \HII\ region lines, the electron densities $n_{\rm e}$ and the effective temperature of the ionizing stars can be determined.  The fluxes and associated uncertainties for these seven far-infrared fine structure line measurements are given in Table~4 with the resolved and unresolved subsets noted in the table.  

\subsubsection{\CII~158\m}
The C$^+$ fine structure transition at 157.74\m\ is the dominant coolant of the neutral interstellar medium and traces PDRs.  Because of the low ionization potential of neutral carbon (11.26~eV), \CII~158\m\ will emanate from neutral surface layers of far-ultraviolet illuminated neutral gas clouds.  C$^+$ is also easy to excite ($\Delta E/k \sim 91$~K and $n_{\rm crit} \sim 3 \cdot 10^3$~cm$^{-3}$) and therefore cools the warm, neutral gas (Tielens \& Hollenbach 1985; Wolfire et al. 1990).  In addition to PDRs, significant contributions to \CII~158\m\ emission can arise from ionized gas in diffuse \HI\ and \HII\ regions, although it is unclear how much \CII~158\m\ comes from these regions (Madden et al. 1993, 1997; Petuchowski \& Bennett 1993; Heiles 1994; Sauty, Gerin \& Casoli 1998; Malhotra et al. 2001; Contursi et al. 2002; Sauvage, Tuffs, \& Popescu 2005).  \CII~158\ is detected in 153 unresolved galaxies and 3$\sigma$ upper limits are determined in another 17 galaxies in the unresolved subset of galaxies.  \CII~158\m\ is detected in all 46 galaxies in the resolved sample.

\subsubsection{\OI~63\m\ \& OI~145\m}
Neutral oxygen has two fine structure transitions at 63\m\ and 145\m\ and has an ionization potential of 13.62~eV.  Atomic oxygen is only found in neutral regions and exists deeper into clouds than C$^+$.  \OI~63\m\ becomes the main coolant in warmer and denser environments ($T > 200$~K and $n > 10^5$~cm$^{-3}$) due to its higher excitation energies and critical densities ($\Delta E/k \sim 228$~K; $n_{\rm crit} \sim 8.5 \cdot 10^5$~cm$^{-3}$ for \OI~63\m\ at $T~\sim 100$~K and $\Delta E/k \sim 325$~K; $n_{\rm crit} \sim 1 \cdot 10^5$~cm$^{-3}$ for \OI~145\m\ at $T~\sim 100$~K).  The \OI~63\m\ line may be particularly strong in the X-ray dissociated regions surrounding active galactic nuclei (Maloney, Hollenbach, \& Tielens 1996; Dale et al. 2004).  In this sample, the \OI\ transitions are always observed in emission except in the ultraluminous infrared galaxy Arp~220, where the 63\m\ line is observed in absorption.  The case of Arp~220 is discussed as part of a progression of emission and absorption line characteristics in a spectroscopic survey of infrared bright galaxies (Fischer et al. 1999) and its absorption spectrum is discussed in detail in Gonz\'alez-Alfonso et al. (2004).  In the unresolved galaxy sample, the \OI~63\m\ line is detected in 93 galaxies with 3$\sigma$ upper limits available for an additional 25, while in the resolved galaxy sample it is detected in 28 galaxies with  3$\sigma$ upper limits available for an additional three galaxies.  In the unresolved galaxy sample, the much fainter \OI~145\m\ line is detected in 20 galaxies and 3$\sigma$ upper limits are measured for another 15.  In the resolved galaxy sample, the \OI~145\m\ line is detected in nine galaxies and 3$\sigma$ upper limits are available for another two resolved systems.

\subsubsection{\NII~122\m}
Neutral nitrogen has an ionization potential of 14.53~eV.  One of the brighter lines of singly-ionized nitrogen, the \NII~122\m\ line has a critical electron density of $3.1 \cdot 10^2$~cm$^{-3}$.  The \NII~122\m\ transition arises only in diffuse, ionized \HII\ regions.  The {\it Cosmic Background Explorer} ({\it COBE}) and the {\it Kuiper Airborne Observatory} provided the first astronomical detections of the \NII~122\m\ line (Wright et al. 1991; Colgan et al. 1993).  This sample greatly expands the number of extragalactic \NII\ detections from earlier studies (Malhotra et al. 2001).  The \NII~122\m\ line is detected in 38 unresolved galaxies and 3$\sigma$ upper limits are measured for an additional 41 of the galaxies in the unresolved subset.  For the resolved subset of galaxies, \NII\ is detected in 16 galaxies and 3$\sigma$ upper limits are reported for another two.   

\subsubsection{\OIII~52\m\ \& 88\m}
An ionizing energy of 35.12~eV is required to create O$^{++}$ from singly-ionized oxygen.  Due to this high ionization potential, the \OIII~52\m\ and 88\m\ transitions occur in \HII\ regions.  Using the methodology of Rubin et al. (1994), the ratio of these two lines, \OIII~88\m/\OIII~52\m, can be used to derive the average electron density $n_{\rm e}$ of these regions within galaxies (Duffy et al. 1987; Carral et al. 1994; Lord et al. 1996; Fischer et al. 1996; Colbert et al. 1999; Unger et al. 2000; Malhotra et al. 2001; Hunter et al 2001).  The \OIII~88\m\ line is detected in 52 galaxies and 3$\sigma$ upper limits are available for 14 galaxies in the unresolved subset.  The \OIII~88\m\ line is detected in 16 galaxies in the resolved galaxy sample while 3$\sigma$ upper limits are available for an additional galaxy in the resolved subset.  The lower signal-to-noise \OIII~52\m\ line is detected in 11 unresolved galaxies with 3$\sigma$ upper limits for another 22 available, whereas the \OIII~52\m\ line is detected in five resolved galaxies and there are an additional six 3$\sigma$ upper limits for the resolved galaxy subset.

\subsubsection{\NIII~57\m}
N$^+$ has a high ionization potential of 47.45~eV and therefore, the 57\m\ transition of N$^{++}$ is only found in the ionized \HII\ regions of the galaxies in this sample.  The ratio \NIII~57\m/\NII~122\m\ provides a measure of the effective temperature $T_{\rm eff}$ (Rubin et al. 1994).  The LWS allowed a more detailed study of the faint \NIII~57\m\ line since there are few detections of this line in the literature (Duffy et al. 1987; Malhotra et al. 2001).  For the unresolved subset of galaxies, the \NIII~57\m\ line is detected in nine galaxies and 3$\sigma$ upper limits are available for another 24 galaxies.  The \NIII~57\m\ line is detected in two resolved galaxies and 3$\sigma$ upper limits are determined for six additional galaxies resolved by the LWS aperture.

\subsection{Molecular Lines}
Molecular line fluxes are reported in Table~5 for several of the brightest galaxies observed in this sample with {\it IRAS} 60\m\ fluxes typically higher than 100~Jy.  Molecular lines are observed in both emission and absorption for these galaxies.  

\subsubsection{H$_2$O}
Water (H$_2$O) has been reported in two galaxies in this sample.  The unresolved 101\m\ ortho-para-H$_2$O pair is found in absorption for NGC~4945.  All five H$_2$O transitions reported in this paper are observed in absorption in Arp~220.  Detailed analysis of the far-infrared H$_2$O lines in Arp~220 and Mrk~231 are presented in Gonz\'alez-Alfonso et al. (2004, 2008).
 
\subsubsection{OH}
Six transitions of Hydroxyl (OH) have been detected in this sample. The OH 53\m\ transition is measured in absorption and the 163\m\ transition in emission for NGC~253 and Arp~220.  All galaxies in Table~5 show OH from the ground level at 119\m, and with the exception of the archetypical Seyfert 2 galaxy NGC~1068, this transition is measured in absorption.  In fact, the OH lines detected for NGC~1068, at 79, 119, and 163\m, are observed in emission, suggesting a unique excitation environment (Spinoglio et al. 2005).  The ultraluminous Seyfert~1 galaxy Mrk~231 shows an absorption line spectrum very similar to that of the ultraluminous galaxy Arp~220.  The multiple OH detections in NGC~253, NGC~1068, Mrk~231, and the megamaser galaxies IRAS~20100-4156 and III~Zw~35 have been previously reported and analyzed in detail (Bradford et al. 1999; Kegel et al. 1999; Goicoechea, Mart\'in-Pintado, \& Cernicharo 2005; Spinoglio et al. 2005; Gonz\'alez-Alfonso et al. 2008).  In Arp~220, all six transitions of OH reported in this paper are detected.  A detailed analysis of the far-infrared absorption spectrum of Arp~220 and the implications for understanding the ``\CII~158\m\ deficit'' (see \S~6.1, Luhman et al. 2003) is discussed in  Gonz\'alez-Alfonso et al. (2004).

\subsection{Unidentified Line}
An unidentified emission line at 74.24\m, also reported in NGC~7027 (Liu et al. 1996) and RWC~103 (Oliva et al. 1999), is detected in NGC~1068.  The line flux and associated uncertainty for this line is listed in Table~5.

\subsection{Serendipitous Galactic \CII~158\m}
The Far-Infrared Absolute Spectrophotometer (FIRAS) on the {\it COBE} satellite conducted an unbiased survey of the far-infrared emission from our Galaxy.  The FIRAS spectral line survey included the emission lines from \CII~158\m, \NII~122\m\ and 205\m, \CI~370\m\ and 609\m, and CO~$J=2-1$ through $J=5-4$ with a resolution of 7\degr\ and were first reported by Wright et al. (1991).  The \CII~158\m\ line had sufficient strength to be mapped by FIRAS, and Bennett et al. (1994) present detailed maps of this emission line.  The all-sky maps of the \CII~158\m\ line show the highest concentration at low Galactic latitudes ($|b| < 20$\degr).  The cosecant relation provided by Bennett et al. (1994) for Galactic \CII~158\m\ emission based on {\it COBE} data is
\be
I([{\rm C~II}]~158\micron) = (1.43 \pm 0.12) \cdot 10^{-6} \csc |b|~{\rm ergs}~{\rm cm}^{-2}~{\rm s}^{-1}~{\rm sr}^{-1},
\label{eq:CII}
\ee
nominally applicable for $|b|>15$\degr.

There are approximately 40 galaxies in this sample that reside at low Galactic latitudes of $|b| < 20$\degr.  Galactic \CII~158\m\ contamination for low Galactic latitude galaxies can be a concern, particularly if they have recessional velocities smaller than the velocity resolution of the LWS ($|v| < 1500$~km~s$^{-1}$).  The impact of Galactic \CII~158\m\ contamination in such systems can be directly addressed via off-galaxy/sky observations made in concert with the targeted extragalactic observations.  Table~6 lists six galaxies with $|v| < 1500$~km~s$^{-1}$ for which sky observations are available.  These off-galaxy positions were typically carried out 4\arcmin--6\arcmin\ away from the targeted extragalactic direction.  The level of Galactic \CII~158\m\ contamination is between 10\% and 25\% of the total \CII~158\m\ in the LWS aperture for these six galaxies.  The \CII~158\m\ line fluxes listed in Table~4 for these six galaxies have had the foreground Milky Way \CII~158\m\ from Table~6 removed.  

Equation~\ref{eq:CII} provides another method for estimating the Galactic \CII~158\m\ contamination for galaxies with $|v| < 1500$~km~s$^{-1}$.  To enhance the comparison between LWS observations and predictions from the {\it COBE} relation, an additional six galaxies have been added to Table~6, galaxies with recessional velocities large enough such that Milky Way \CII~158\m\ contamination is not an issue.  The maps at 100\m\ and the LWS line spectra containing both Galactic C$^+$ and extragalactic C$^+$ for three higher redshift galaxies are presented in Figure~\ref{fig:CII_MW}.  The C$^+$ line associated with each target galaxy is located at the redshift of the galaxy, whereas the foreground Milky Way C$^+$ is at the rest wavelength of 157.74\m.  The \CII~158\m\ intensities predicted from Equation~\ref{eq:CII} agree with the observations to within a factor of 2 for all galaxies in Table~6 except Maffei~2, which lies a half degree from the Galactic plane.  For the remaining dozen or so galaxies with $|v| < 1500$~km~s$^{-1}$ and $|b| < 20$\degr\ that lack off-galaxy/sky observations, the contamination from Milky Way \CII~158\m\ is predicted from Equation~\ref{eq:CII} to be no more than 25\%.  For the 55 higher latitude ($|b| > 20$\degr) galaxies with $|v| < 1500$~km~s$^{-1}$ and no off-source observation, any Milky Way \CII~158\m\ contamination is likely much less than 10\%.       

The four high latitude detections of Galactic \CII~158\m\ listed in Table~6 may either be due to 
the warm ionized medium (Petuchowski \& Bennett 1993) or high latitude molecular clouds (Magnani et al. 1996).  The {\it IRAS} Sky Survey Atlas images at 60\m\ and 100\m\ reveal extended Galactic emission in the same direction as the four high latitude \CII~158\m\ detections.  Reach et al. (1998) describe the location of the UGCA~332 observation as a high latitude warm infrared excess \HII\ region around the nearby B star Spica using far-infrared (60--240\m) data from the {\it COBE} Diffuse Infrared Background Experiment and the Leiden-Dwingeloo \HI\ survey (Hartmann \& Burton 1997), and the spectrum for this object in Figure~\ref{fig:CII_MW} confirms the foreground nature of this line emission.
              
\section{Statistical Trends in the Line Data}
In this section the trends in the far-infrared fine structure line fluxes are examined for the subset of 181 galaxies unresolved by the LWS beam.  Line-to-line and line-to-far-infrared ratios are examined across a broad range of 60\m/100\m\ and {\it FIR/B} values.  The 60\m/100\m\ ratio is an indicator of the dust heating intensity in galaxies, which is related to the star formation activity in a galaxy.  The {\it FIR/B} ratio compares the luminosity reprocessed by dust to that of escaping starlight, indicating star formation activity along with the effects of extinction.  The intent here is to identify major trends or lack thereof, as opposed to carrying out a detailed physical analysis with model comparisons.

\subsection{\CII~158\m/{\it FIR}}
\begin{figure}
 \plotone{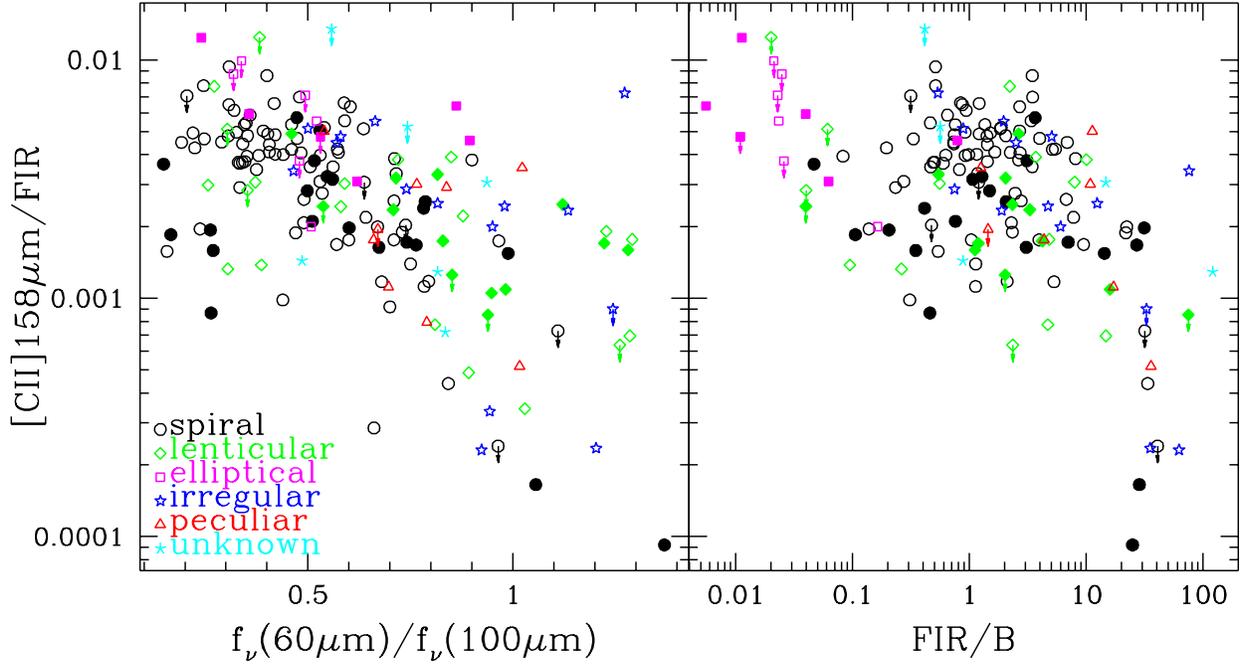}
 \caption{The ratio of \CII~158\m\ to far-infrared continuum is plotted against the {\it IRAS} 60\m/100\m\ and {\it FIR/B} ratios for galaxies unresolved by the LWS.  Galaxies of spiral ($T=0-9$), lenticular ($T=-3,-2,-1$), elliptical ($T=-6,-5,-4$), irregular ($T=10,11,90$), peculiar ($T=99$), and unknown morphology are respectively plotted as circles, diamonds, squares, stars, and asterisks.  AGN are indicated by filled symbols.  Regardless of morphology, the \CII~158\m/{\it FIR} ratio decreases as the 60\m/100\m\ and {\it FIR/B} ratios increase.}
\label{fig:CII_FIR}
\end{figure}

Previous studies (Malhotra et al. 1997, 2001; Luhman et al. 1998, 2003; Leech et al. 1999; Negishi et al. 2001) of the \CII~158\m/{\it FIR} ratio for galaxies reveal a trend with dust heating intensity as measured by 60\m/100\m\ and/or {\it FIR/B}.  Figure~\ref{fig:CII_FIR} shows this trend for the 181 unresolved galaxies, a trend that broadly holds for all morphological types.  A \CII/{\it FIR} ratio that decreases from 1\% to 0.1\% with increasingly warm infrared color is typical of normal and starburst galaxies, confirming earlier studies based on smaller samples (Crawford et al. 1985; Stacey et al. 1991; Malhotra et al. 1997, 2001; Luhman et al. 2003; Verma et al. 2005).  
As the dust temperature increases for the most actively star-forming galaxies in this sample (60\m/100\m\ $\geq 0.8$), the \CII/{\it FIR} ratio reaches levels as low as 0.01\%.  The elliptical galaxies NGC~6958 and NGC~1052 are a factor of 2--5 lower than typical values of \CII/{\it FIR} for normal galaxies as first reported by Malhotra et al. (2000).  However, the remaining early-type galaxies (ellipticals and lenticulars) appear to have \CII/{\it FIR} ratios similar to those of the other morphological types.  \CII~158\m\ emission in irregular galaxies is higher relative to the \CII~158\m\ emission in spiral galaxies of the same far-infrared color temperature shown in Figure~\ref{fig:CII_FIR}, as was also noted by Hunter et al. (2001).  
                                 
There is a large spread in the \CII/{\it FIR} ratio for a given 60\m/100\m\ and {\it FIR/B} ratio.  Despite this, there are several observed trends in the \CII/{\it FIR} shown in Figure~\ref{fig:CII_FIR} and several possible explanations for these trends in galaxies.
 
1. The \CII/{\it FIR} peaks for normal, star-forming galaxies with a 60\m/100\m\ ratio of 0.3--0.6 and {\it FIR/B} ratio between 0.01 and 1, consistent with the earlier studies mentioned above.  These galaxies may have a higher fraction of intermediate-mass stars that are efficient at producing ultraviolet and \CII~158\m\ line emission, thus, causing the peak for these galaxies.  Alternatively, the high \CII/{\it FIR} values for at least the early-types may simply due to a dearth of far-infrared emission.
        
2. The \CII/{\it FIR} ratios for quiescent galaxies below a 60\m/100\m\ ratio of 0.3 and {\it FIR/B} ratio of 0.1 are, on average, similar to or slightly less than those for normal, star-forming galaxies.  There is some evidence for lower \CII/{\it FIR} ratios in the quiescent galaxies of the Virgo Cluster as suggested in this study and Leech et al. (1999).  Quiescent galaxies with this range of 60\m/100\m\ and {\it FIR/B} have a larger old low-mass stellar population than normal galaxies, and produce less ultraviolet and \CII~158\m\ line emission, possibly causing a decrease in the observed \CII/{\it FIR} ratio.

3. The \CII/{\it FIR} decreases with increasing 60\m/100\m\ and {\it FIR/B} ratio (Malhotra et al. 1997, 2001).  This trend has been seen within our Galaxy (Nakagawa et al. 1995; Bennett et al. 1994) and is not surprising for a sample of galaxies spanning a large range of 60\m/100\m\ and {\it FIR/B} ratios.  Galaxies with 60\m/100\m\ ratios greater than 0.6 and {\it FIR/B} ratios greater than unity have increasingly warmer dust temperatures, most likely due to more extreme star formation.  These actively star-forming galaxies have a large proportion of massive O stars that produce hard ultraviolet radiation.  Several explanations from previous studies have been offered for the decrease in the \CII/{\it FIR} ratio in galaxies with the warmest dust temperatures.  Malhotra et al. (2001) propose that the decrease in \CII/{\it FIR} is due to the dust grains becoming more positively charged and less efficient at heating the gas for high ratios of ultraviolet flux-to-gas density ($G_0/n$) according to PDR models.  Negishi et al. (2001) attribute this decrease in \CII/{\it FIR} to either an increase in the collisional de-excitation of the \CII~158\m\ transition at high densities or a decrease in the ionized component of the \CII~158\m\ emission.  For a sample of 15 ultraluminous infrared galaxies Luhman et al. (2003) report a deficiency of \CII~158\m, consistent with the decrease in the \CII/{\it FIR} ratio at high 60\m/100\m\ ratios and explain this deficiency as the result of 
non-PDR contributions to the far-infrared continuum, possibly from dust-bounded ionized regions.
     
\subsection{\OI~63\m/{\it FIR}}
The \OI~63\m/{\it FIR} ratio for galaxies shows no trend with 60\m/100\m\ and a decreasing trend with {\it FIR/B} as displayed in Figure~\ref{fig:OI_FIR} for the 181 galaxies of the unresolved subset, plotted according to morphological type.  An \OI~63\m/{\it FIR} of 0.05\%--1\% characterizes these galaxies.  Although \CII~158\m/{\it FIR} tends to decrease with increasing 60\m/100\m, the same is not found for \OI~63\m/{\it FIR}, consistent with earlier studies by Malhotra et al. (2001) and Negishi et al. (2001).  Therefore, as the heating environment in galaxies gets warmer, \CII~158\m\ becomes less dominant while \OI~63\m\ becomes more important in the cooling of the interstellar medium (see Section 6.5).  

\subsection{\NII~122\m/{\it FIR}}
\begin{figure}
 \plotone{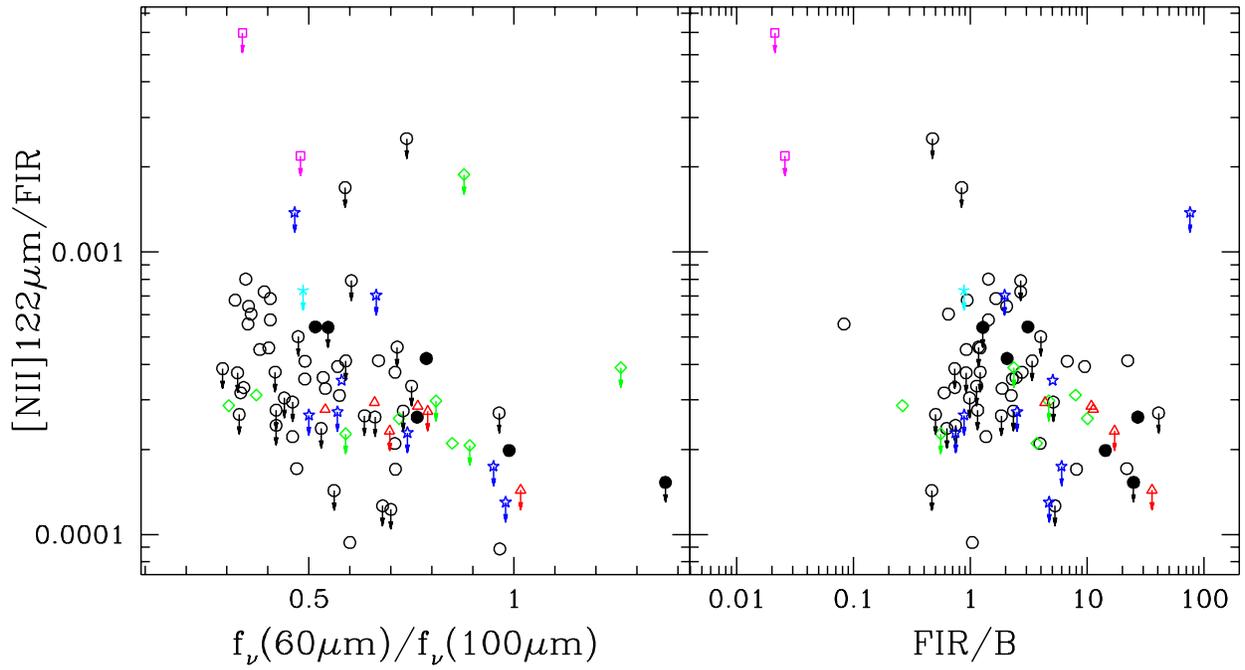}
 \caption{The ratio of \NII~122\m\ to far-infrared continuum is plotted against the {\it IRAS} 60\m/100\m\ and {\it FIR/B} ratios for galaxies unresolved by the LWS.  The \NII~122\m/{\it FIR} ratio decreases as the 60\m/100\m\ and {\it FIR/B} ratios increase, similar to the behavior of \CII~158\m/{\it FIR} line.  The symbols are the same as those in Figure~\ref{fig:CII_FIR}.}
\label{fig:NII_FIR}
\end{figure}

The \NII~122\m/{\it FIR} ratio for galaxies reveals a trend with dust temperature as measured by the 60\m/100\m\ ratio (Malhotra et al. 2001).  The \NII~122\m/{\it FIR} for the 181 galaxies of the unresolved subset in this sample, plotted according to morphological type, is presented in Figure~\ref{fig:NII_FIR}.  A \NII~122\m/{\it FIR} of 0.01\%--0.1\% characterizes these galaxies.  The \NII~122\m/{\it FIR} follows a similar decreasing trend as \CII~158\m/{\it FIR} with 60\m/100\m\ and {\it FIR/B} as suggested in Figure~\ref{fig:NII_FIR}.  There is a clear decrease in \NII~122\m/{\it FIR} as 60\m/100\m\ and {\it FIR/B} increase for spirals and irregulars.  For morphologies other than spirals and irregulars, there are few detections of \NII~122\m\ and no trend is discernible.  From {\it COBE} observations of the Milky Way, a correlation between \CII~158\m\ and \NII~205\m\ was found (Bennett et al. 1994), therefore, it is not surprising that \NII~122\m\ and \CII~158\m\ in galaxies exhibit some of the same characteristics over a broad range of heating environments.

\subsection{\OIII~88\m/{\it FIR}}
\begin{figure}
 \plotone{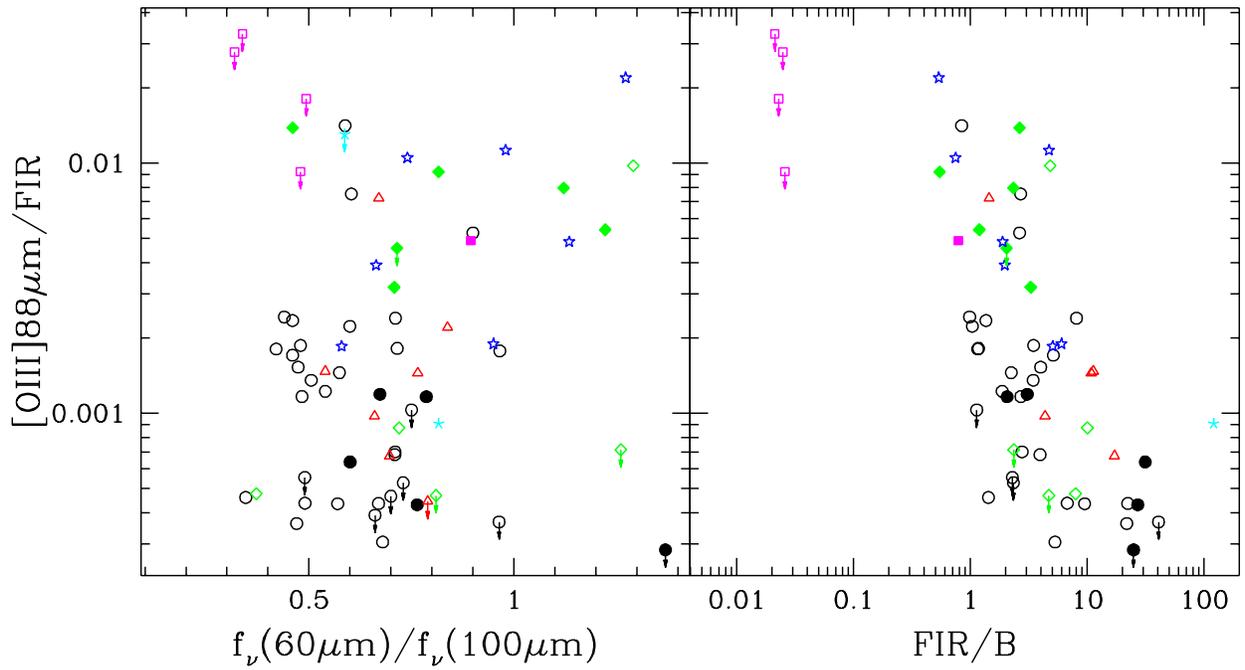}
 \caption{The ratio of \OIII~88\m\ to far-infrared continuum is plotted against the {\it IRAS} 60\m/100\m\ and {\it FIR/B} ratios for galaxies unresolved by the LWS.  The \OIII~88\m/{\it FIR} ratio increases with increasing 60\m/100\m\ ratio and decreases with increasing {\it FIR/B} ratio.  The symbols are the same as those in Figure~\ref{fig:CII_FIR}.}
\label{fig:OIII_FIR}
\end{figure}

The \OIII~88\m/{\it FIR} ratio for the galaxies in the unresolved subset is presented in Figure~\ref{fig:OIII_FIR}.  An \OIII~88\m/{\it FIR} of 0.03\%--2\% characterizes this sample.  While there is a large scatter among the data and a small number of \OIII~88\m\ detections above 60\m/100\m\ ratios of 0.9, there seems to be a weak increasing trend in \OIII~88\m/{\it FIR} as the 60\m/100\m\ ratio increases in Figure~\ref{fig:OIII_FIR}a.  On average, there is relatively more \OIII~88\m\ emission in warmer galaxies, presumably due to a higher density of \HII\ regions in these galaxies.  This overall increase in the \OIII~88\m/{\it FIR} ratio is also noted in Negishi et al. (2001) for a smaller set of galaxy observations.  In Figure~\ref{fig:OIII_FIR}b, the \OIII~88\m/{\it FIR} appears to decrease as the {\it FIR/B} ratio increases.  Malhotra et al. (2001) point out this anticorrelation between \OIII~88\m/{\it FIR} and {\it FIR/B}, but they attribute the effect to the observations of two irregular galaxies in their sample.  In this study, many new observations of the \OIII~88\m\ line are included, and a decreasing trend in \OIII~88\m/{\it FIR} with increasing {\it FIR/B} is discovered. 

\subsection{\OI~63\m/\CII~158\m}
\begin{figure}
 \plotone{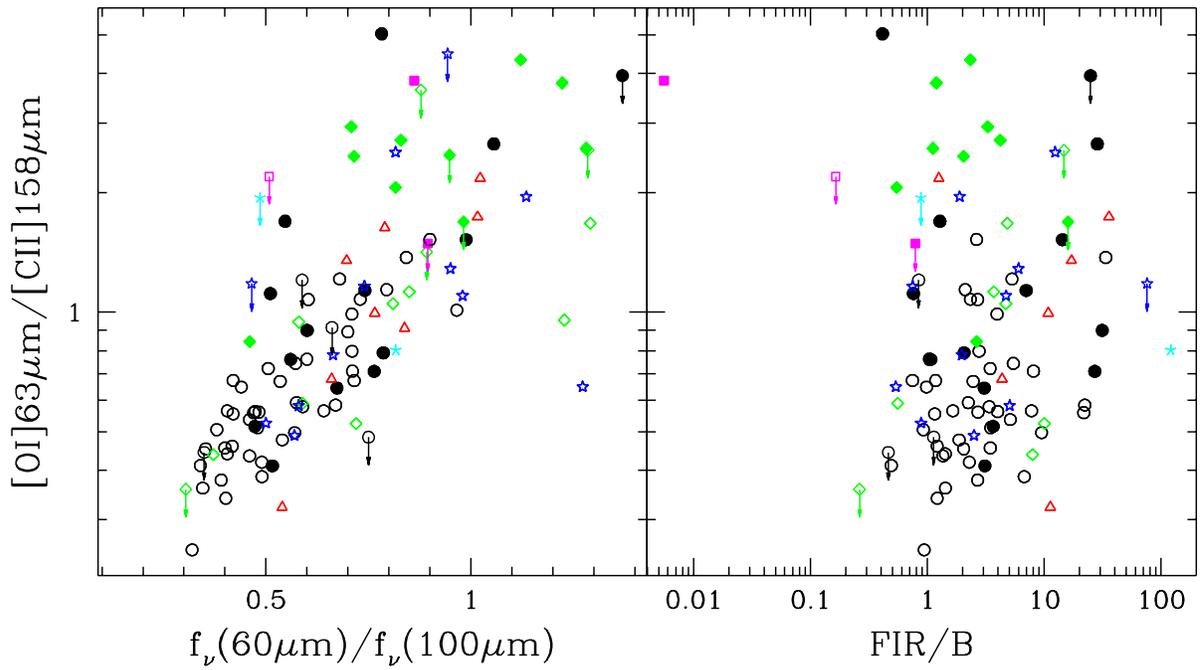}
 \caption{The ratio of \OI~63\m/\CII~158\m\ is plotted against the {\it IRAS} 60\m/100\m\ and {\it FIR/B} ratios for galaxies unresolved by the LWS.  The \OI~63\m/\CII~158\m\ ratio increases as the 60\m/100\m\ ratio increases, but shows no correlation with the {\it FIR/B} ratio.  The symbols are the same as those in Figure~\ref{fig:CII_FIR}.}
\label{fig:OI_CII_vs_60_100}
\end{figure}

In Figures~\ref{fig:CII_FIR} and \ref{fig:OI_FIR}, \CII~158\m/{\it FIR} is shown to decrease with increasing 60\m/100\m\ and {\it FIR/B} while \OI~63\m/{\it FIR} remained steady with 60\m/100\m\ and {\it FIR/B}.  In Figure~\ref{fig:OI_CII_vs_60_100}, the \OI~63\m/\CII~158\m\ ratio is plotted against 60\m/100\m\ and {\it FIR/B}, and a rise in \OI~63\m/\CII~158\m\ ratio is found as 60\m/100\m\ increases for all morphologies but no conclusive trend in the \OI~63\m/\CII~158\m\ ratio is found as {\it FIR/B} increases.  From Figure~\ref{fig:OI_CII_vs_60_100}a, \OI~63\m\ begins to dominate cooling in the interstellar medium of warmer galaxies (60\m/100\m\ $\geq 0.8$), consistent with the results reported by Malhotra et al. (2001).    

\subsection{\NII~122\m/\CII~158\m}
\begin{figure}
 \plotone{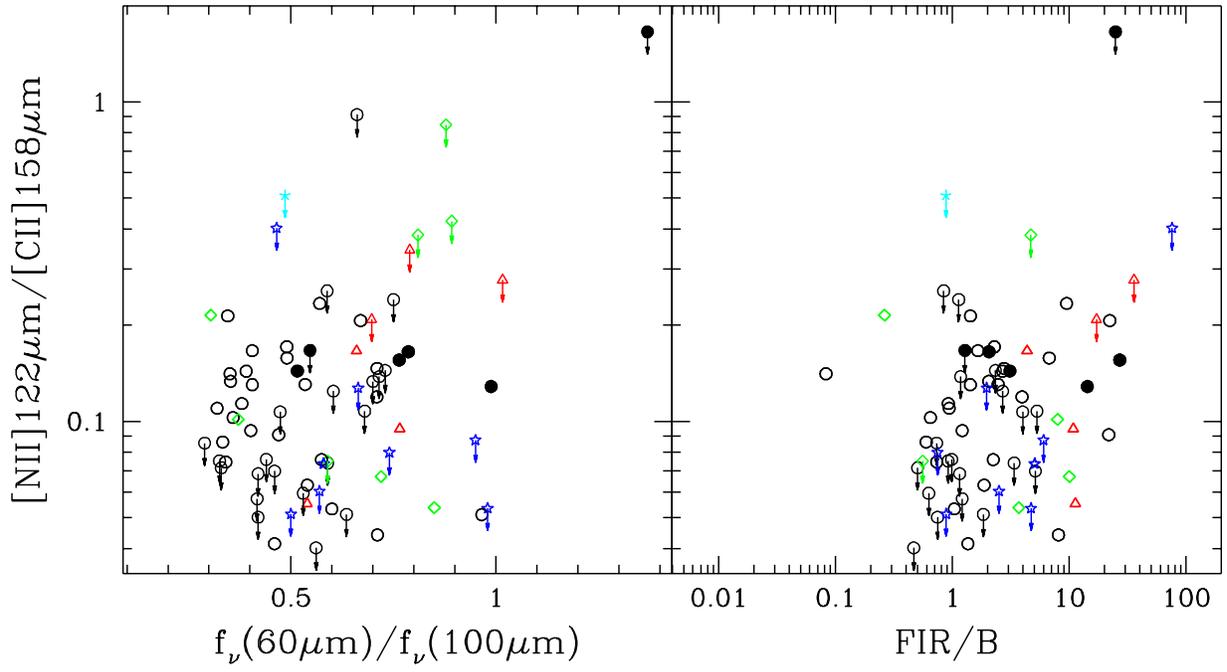}
 \caption{The ratio of \NII~122\m/\CII~158\m\ is plotted against the {\it IRAS} 60\m/100\m\ and {\it FIR/B} ratios for galaxies unresolved by the LWS.  The symbols are the same as those in Figure~\ref{fig:CII_FIR}.}
\label{fig:NII_CII}
\end{figure}

The \NII~122\m/\CII~158\m\ ratio remains relatively constant across a broad range of 60\m/100\m\ and {\it FIR/B} for all morphological types as shown in Figure~\ref{fig:NII_CII}.  The median value of \NII~122\m/\CII~158\m\ for this sample is 0.11 when both lines have been detected, consistent with what {\it COBE} observed for the Milky Way (Wright et al. 1991; Bennett et al. 1994), but lower than what models predict if \CII~158\m\ were only produced in \HII\ regions (Rubin 1985).  The similar decreasing behaviors in the \CII~158\m/{\it FIR} and \NII~122\m/{\it FIR} ratios with increasing 60\m/100\m\ and {\it FIR/B} ratios shown in Figures~\ref{fig:CII_FIR} and \ref{fig:NII_FIR} along with Figure~\ref{fig:NII_CII} suggests that a significant fraction of \CII~158\m\ arises from \HII\ regions where N$^+$ originates.
Thus, the \NII~122\m/\CII~158\m\ ratio, on average, is nearly constant across a broad range of heating environments (Malhotra et al. 2001).      

\subsection{\OI~145\m/\OI~63\m}
\begin{figure}
 \plotone{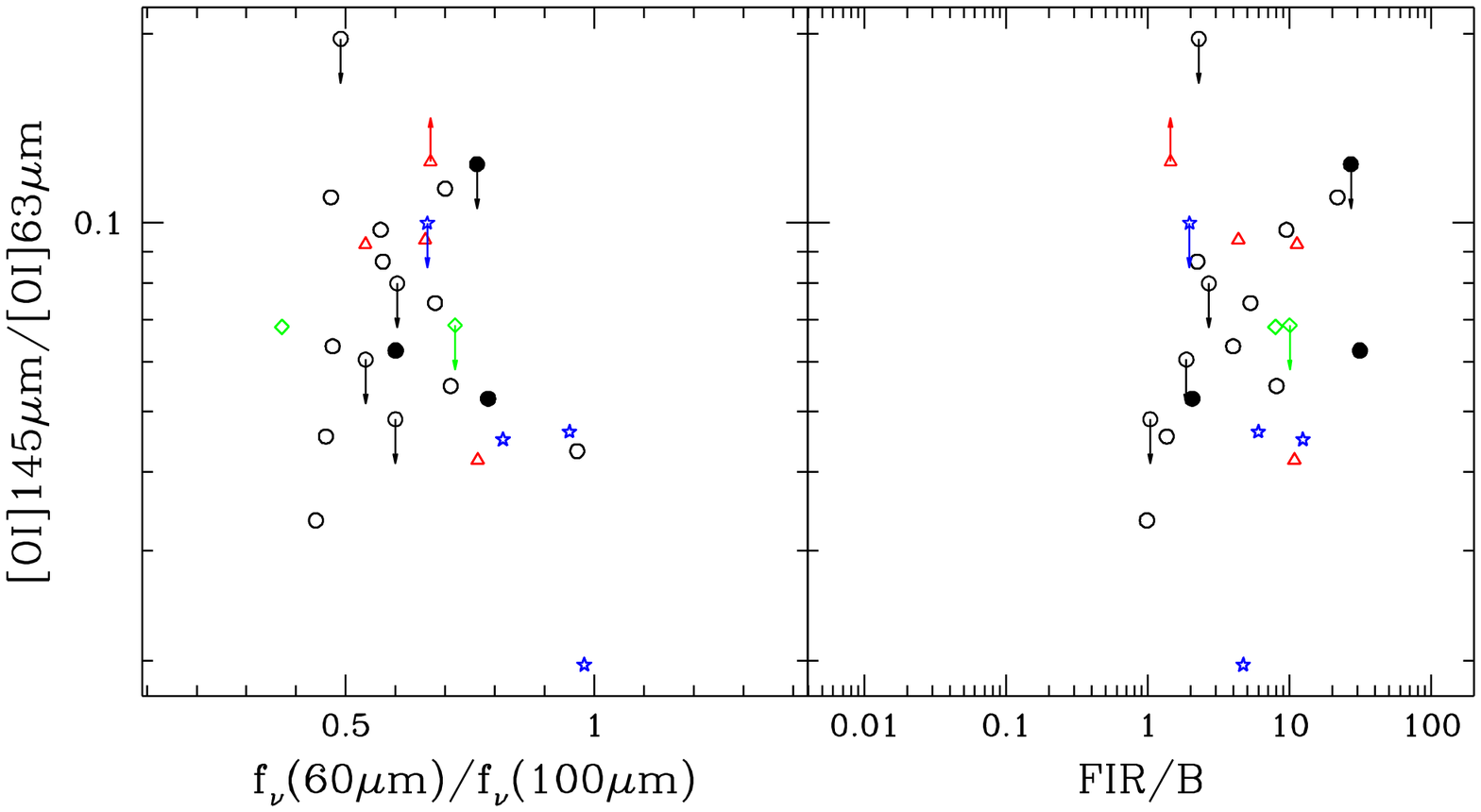}
 \caption{The ratio of \OI~145\m/\OI~63\m\ is plotted against the {\it IRAS} 60\m/100\m\ and {\it FIR/B} ratios for galaxies unresolved by the LWS.  The symbols are the same as those in Figure~\ref{fig:CII_FIR}.}
\label{fig:OI_OI}
\end{figure}

In general, the \OI~63\m\ line goes optically thick for lower column densities than the \OI~145\m\ line does.  The \OI~145\m/\OI~63\m\ measures the gas temperature and the optical depth in the 63\m\ line and rises as the gas temperature increases (Tielens \& Hollenbach 1985; Kaufman et al 1999).  There are few \OI~145\m\ line observations presented in this paper, most of low signal-to-noise.  The low signal-to-noise \OI~145\m\ line has few detections in this sample. Therefore, the \OI~145\m/\OI~63\m\ ratio has no clear trend with the 60\m/100\m\ or {\it FIR/B} ratios as displayed in Figure~\ref{fig:OI_OI}.  Hunter et al. (2001) noted an increase in the \OI~145\m/\OI~63\m\ with increasing {\it FIR/B} ratios between 0.5 and 10 and attributed this increase to an indication of the optical depth effects for \OI~63\m.  

\subsection{(\OI~63\m\ + \CII~158\m) / {\it FIR}}
\begin{figure}
 \plotone{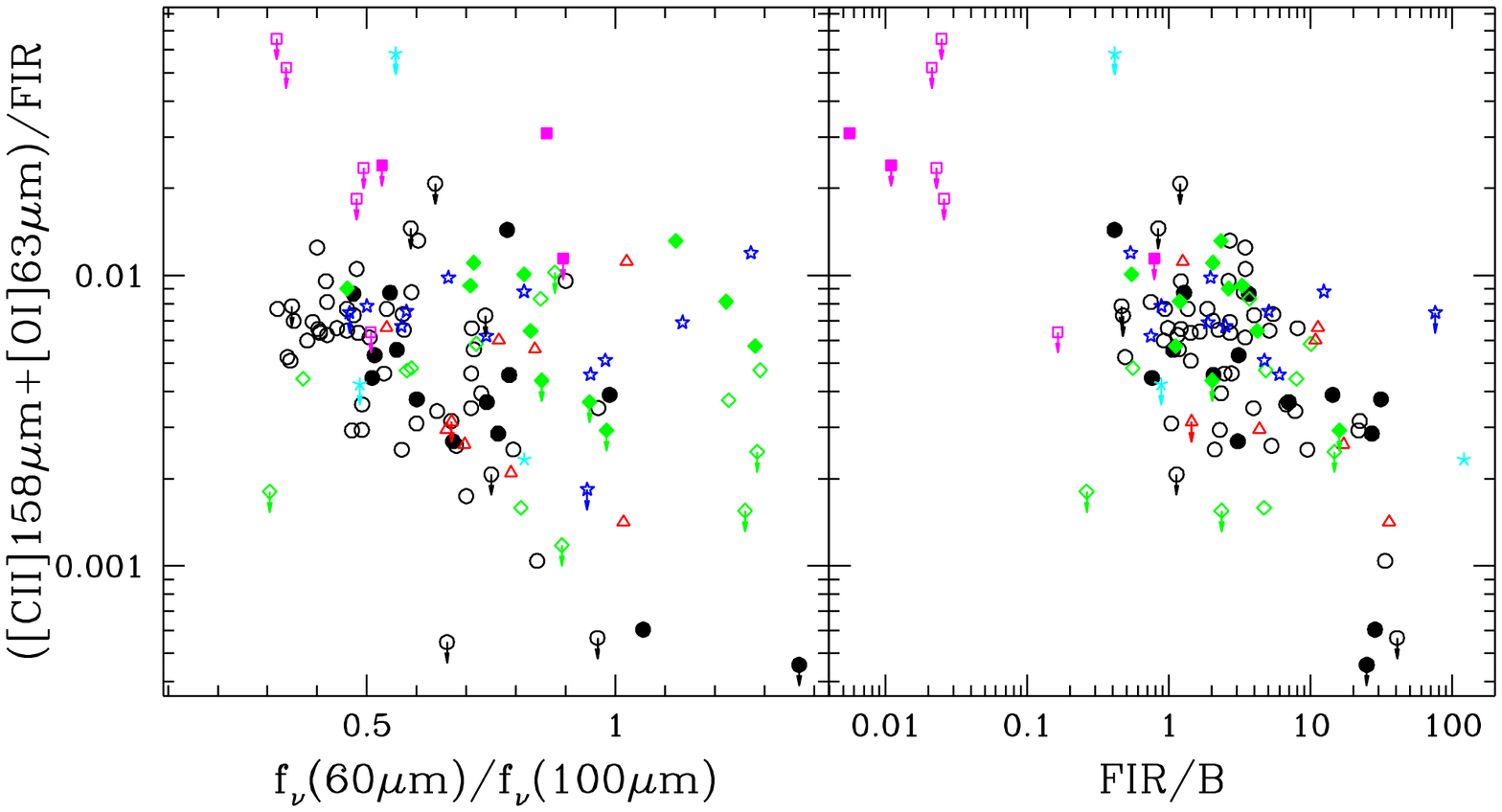}
 \caption{The ratio of (\OI~63\m\ + \CII~158\m)/{\it FIR} is plotted against the {\it IRAS} 60\m/100\m\ and {\it FIR/B} ratios for galaxies unresolved by the LWS.  The symbols are the same as those in Figure~\ref{fig:CII_FIR}.}
\label{fig:OI_CI_FIR}
\end{figure}

The gas heating efficiency of PDRs in galaxies is measured by (\OI~63\m\ + \CII~158\m)/{\it FIR} (Hollenbach \& Tielens 1997) and is plotted against the 60\m/100\m\ ratio and {\it FIR/B} for the unresolved subset of galaxies in the sample discussed in this paper in Figure~\ref{fig:OI_CI_FIR}.  The (\OI~63\m\ + \CII~158\m)/{\it FIR} ratio shows no trend with 60\m/100\m\ and a decreasing trend with {\it FIR/B}.  For a sample of normal galaxies dominated by spirals, Malhotra et al. (2001) noted a decrease in this ratio as 60\m/100\m\ increased.  Although the entire sample of galaxies does not appear to decrease in (\OI~63\m\ + \CII~158\m)/{\it FIR} as the 60\m/100\m\ ratio increases, the spiral galaxies show evidence for a decline in this ratio in warmer dust environments.  The decrease in the (\OI~63\m\ + \CII~158\m)/{\it FIR} with increasing {\it FIR/B} is a result of the decrease in \CII~158\m\ emission shown in Figure~\ref{fig:CII_FIR}.

\subsection{\OIII~88\m/\CII~158\m}
\begin{figure}
 \plotone{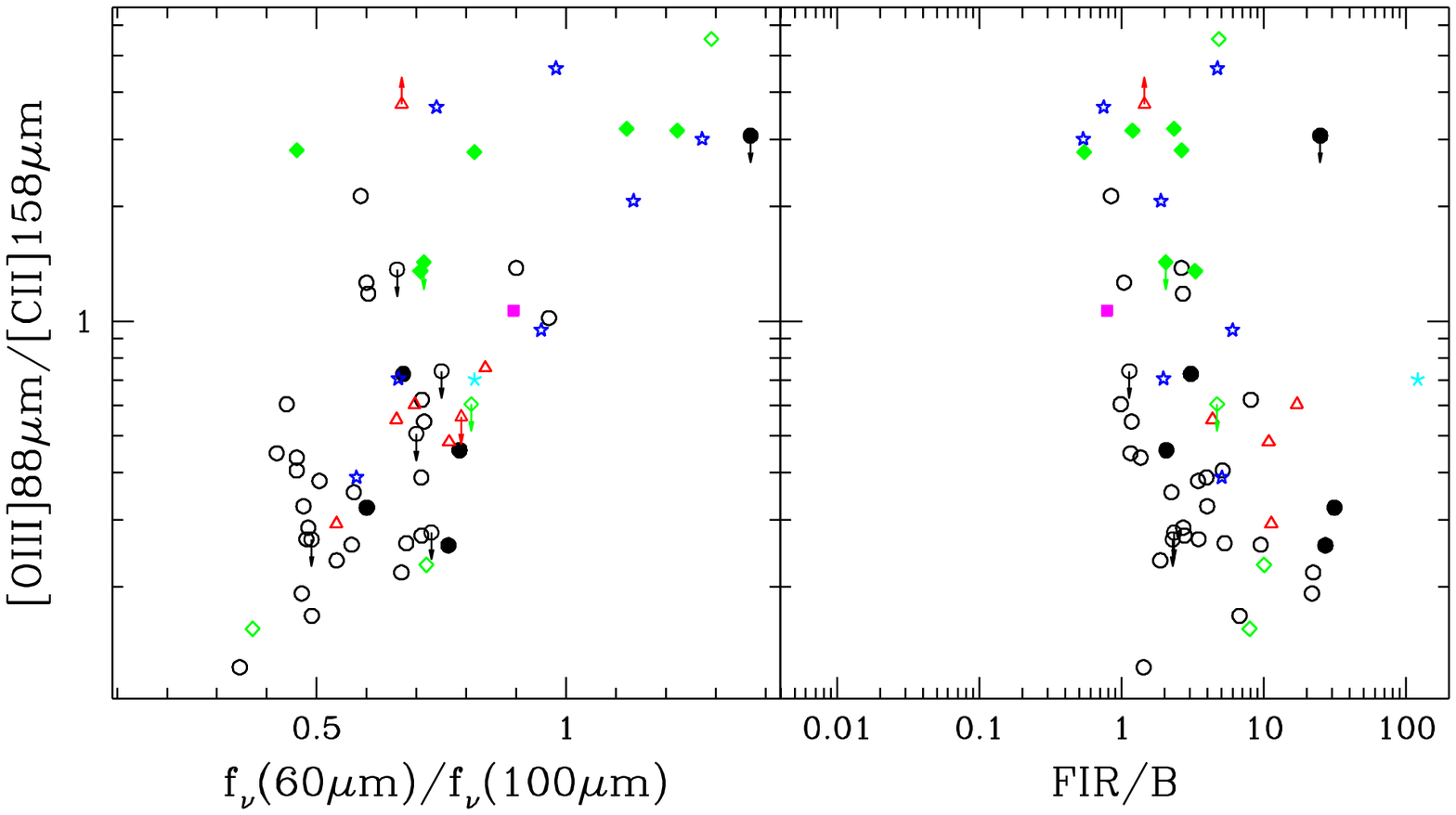}
 \caption{The ratio of \OIII~88\m/\CII~158\m\ is plotted against the {\it IRAS} 60\m/100\m\ and {\it FIR/B} ratios for galaxies unresolved by the LWS.  The \OIII~88\m/\CII~158\m\ ratio increases with increasing 60\m/100mm ratio, but shows no correlation with {\it FIR/B}.  The symbols are the same as those in Figure~\ref{fig:CII_FIR}.}
\label{fig:OIII_CI}
\end{figure}

In Figures~\ref{fig:CII_FIR}a and \ref{fig:OIII_FIR}a, a decrease in \CII~158\m/{\it FIR} and an increase in \OIII~88\m/{\it FIR} emission are shown with increasing 60\m/100\m\ ratio.   Accordingly, the \OIII~88\m/\CII~158\m\ ratio increases with increasing 60\m/100\m\ ratio, as presented in Figure~\ref{fig:OIII_CI}a.  This is an interesting correlation since O$^{++}$ originates in higher density \HII\ regions than does N$^+$.  While the \NII~122\m/\CII~158\m\ ratio did not show a correlation with the 60\m/100\m\ ratio in Figure~\ref{fig:NII_CII}a, the \OIII~88\m/\CII~158\m\ ratio does.  This suggests that the contribution to the C$^+$ emission from \HII\ regions originates in lower density, N$^+$ and O$^+$ regions rather than the highly ionized, more dense O$^{++}$ regions that produce \OIII~88\m.   In comparison, \CII~158\m/{\it FIR} and \OIII~88\m/{\it FIR} are shown to decrease with increasing {\it FIR/B} ratio in Figures~\ref{fig:CII_FIR}b and \ref{fig:OIII_FIR}b.  In Figure~\ref{fig:OIII_CI}b, the \OIII~88\m/\CII~158\m\ ratio is plotted against the {\it FIR/B} ratio.  There is evidence for a decreasing trend in \OIII~88\m/\CII~158\m\ with increasing {\it FIR/B}.  The irregulars, for example, show a decrease in \OIII~88\m/\CII~158\m\ due to a larger decrease in \CII~158\m\ than \OIII~88\m\ emission with increasing {\it FIR/B} ratio.  The irregulars also have higher \OIII~88\m/\CII~158\m\ ratios than most spirals as noted by Hunter et al. (2001) and shown in Figure~\ref{fig:OIII_CI}.  The higher \OIII~88\m/\CII~158\m\ ratios observed in irregulars are likely due to stars with higher effective temperatures found in the \HII\ regions of these galaxies that produce doubly ionized oxygen but little C$^+$.      

\subsection{\OIII~88\m/\OI~63\m}
\begin{figure}
 \plotone{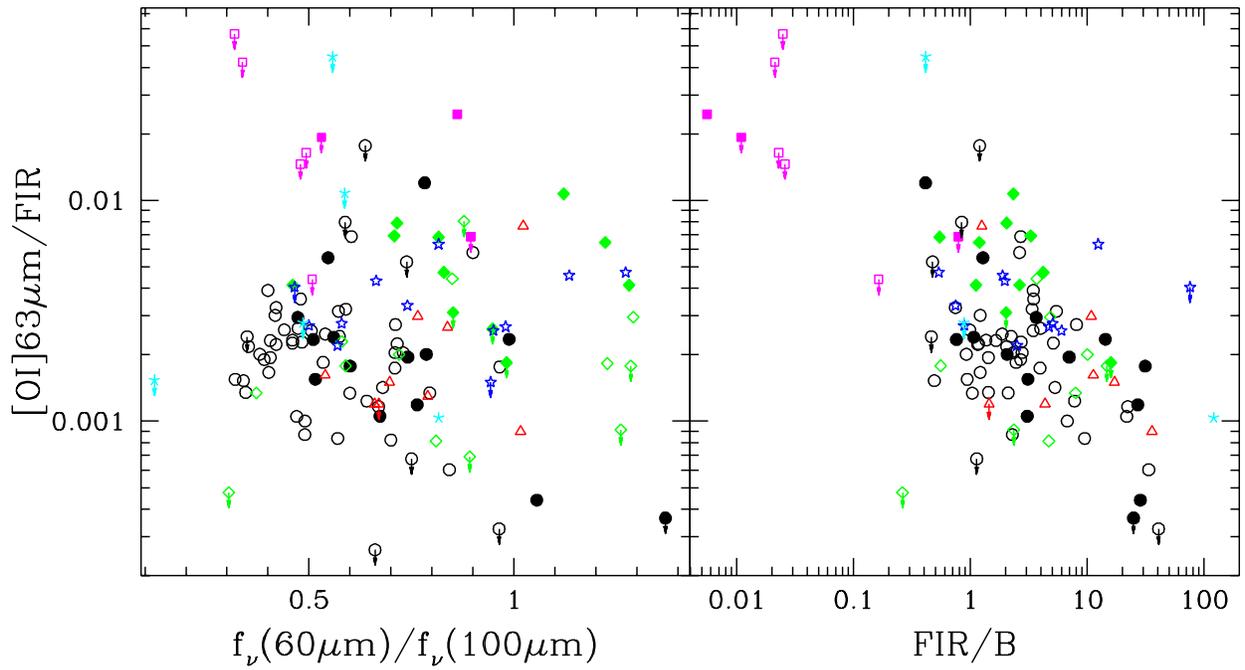}
 \caption{The ratio of \OI~63\m\ to far-infrared continuum is plotted against the {\it IRAS} 60\m/100\m\ and {\it FIR/B} ratios for galaxies unresolved by the LWS.  The \OI~63\m/{\it FIR} shows no trend with either 60\m/100\m\ or {\it FIR/B}.  The symbols are the same as those in Figure~\ref{fig:CII_FIR}.}
\label{fig:OI_FIR}
\end{figure}

\begin{figure}
 \plotone{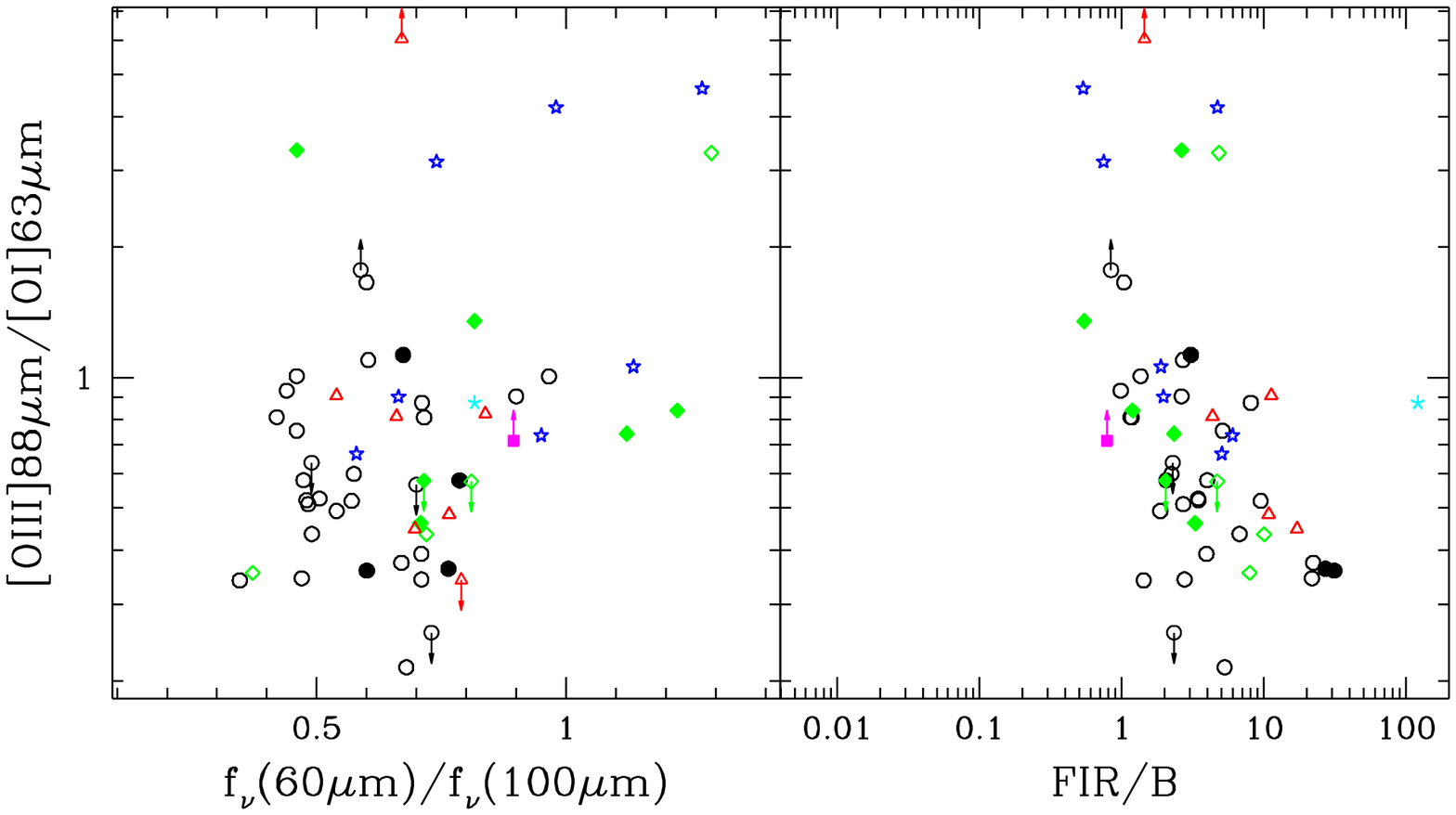}
 \caption{The ratio of \OIII~88\m/\OI~63\m\ is plotted against the {\it IRAS} 60\m/100\m\ and {\it FIR/B} ratios for galaxies unresolved by the LWS.  The symbols are the same as those in Figure~\ref{fig:CII_FIR}.}
\label{fig:OIII_OI}
\end{figure}

In Figure~\ref{fig:OI_FIR}, the \OI~63\m\ emission remains relatively constant when normalized to far-infrared across a broad range of interstellar medium environments measured by the 60\m/100\m\ ratio.  The relative drop seen in \CII~158\m\ emission is not observed for \OI~63\m.  By comparison, \OIII~88\m/{\it FIR} shows a weak, increasing trend with increasing 60\m/100\m\ ratio in Figure~\ref{fig:OIII_FIR}a.  The \OIII~88\m/\OI~63\m\ ratio is plotted against the 60\m/100\m\ ratio in Figure~\ref{fig:OIII_OI}a.  The \OIII~88\m/\OI~63\m\ ratio has no obvious trend that spans all 60\m/100\m\ ratios, unlike \OIII~88\m/\CII~158\m.  In Figures~\ref{fig:OI_FIR}b and \ref{fig:OIII_FIR}b, the \OI~63\m/{\it FIR} ratio shows little trend with {\it FIR/B} while the \OIII~88\m/{\it FIR} ratio falls as the {\it FIR/B} ratio increased.  A slight decline in the \OIII~88\m/\OI~63\m\ ratio is observed as the {\it FIR/B} ratio increases and is shown in Figure~\ref{fig:OIII_OI}b.  Similar to the \OIII~88\m/\CII~158\m\ ratio, the \OIII~88\m/\OI~63\m\ ratio for irregulars decreases noticeably with increasing {\it FIR/B}.  Irregular galaxies also show an elevated \OIII~88\m/\OI~63\m\ ratio when compared to spirals as mentioned by Hunter et al. (2001).     

\subsection{\OIII~88\m/\NII~122\m}
\begin{figure}
 \plotone{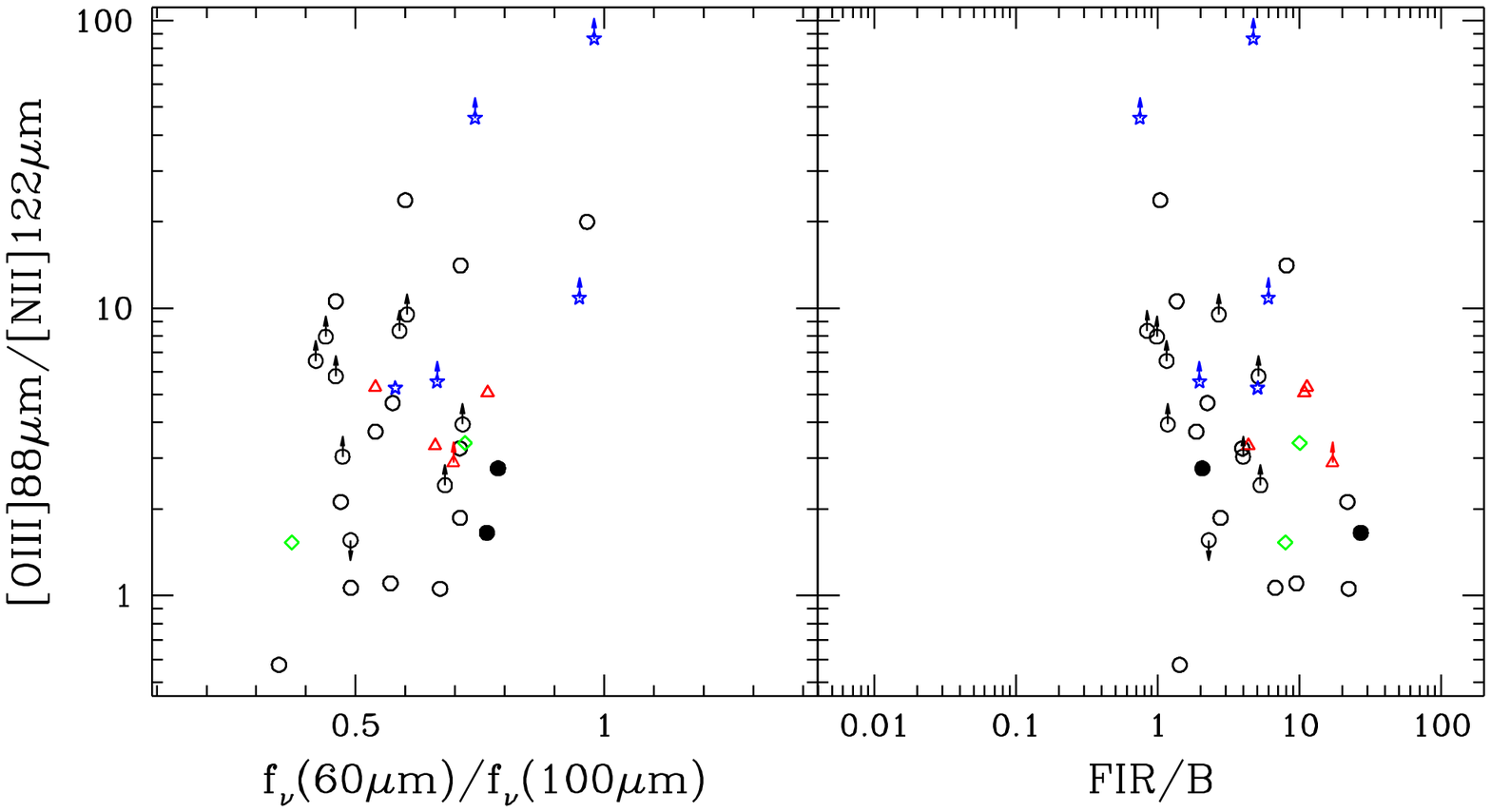}
 \caption{The ratio of \OIII~88\m/\NII~122\m\ is plotted against the {\it IRAS} 60\m/100\m\ and {\it FIR/B} ratios for galaxies unresolved by the LWS.  The symbols are the same as those in Figure~\ref{fig:CII_FIR}.}
\label{fig:OIII_NII}
\end{figure}

\OIII~88\m\ originates in higher density \HII\ regions (excitation potential=35~eV), and \NII~122\m\ originates in lower density \HII\ regions (excitation potential=14.5~eV).  The \OIII~88\m/\NII~122\m\ ratio is plotted against the 60\m/100\m\ and {\it FIR/B} ratios in Figure~\ref{fig:OIII_NII} for galaxies unresolved by the LWS.  Perhaps there are correlations for \OIII~88\m/\NII~122\m\ with the 60\m/100\m\ and {\it FIR/B} ratios, though too few data are available to have confidence in these trends. 

\section {Summary}

ISO LWS far-infrared line and continuum fluxes for a sample of 227 galaxies selected from the {\it ISO} Data Archive spanning an {\it IRAS} 60\m/100\m\ color range of 0.2--1.4 and 60\m\ flux densities between 0.1 Jy and 1300 Jy are presented.  The far-infrared lines detected in this sample include the seven fine structure lines (\CII~158\m, \OI~145\m, \NII~122\m, \OIII~88\m, \OI~63\m, \NIII~57\m, \OIII~52\m) and multiple OH (53\m, 65\m, 79\m, 84\m, 119\m, 163\m) and H$_2$O (59\m, 67\m, 75\m, 101\m, 108\m) transitions.  An unidentified line at 74.24\m\ previously reported in NGC~7027 is detected in NGC~1068.  Serendipitous detections of Milky Way \CII~158\m\ are also observed in twelve sky positions.  This sample is the largest collection of far-infrared line observations ever assembled and includes 465 independent LWS observations yielding some 1300 line fluxes, 600 line flux upper limits, and 800 continuum fluxes.    

The data presented here can be separated into two subsets, one where the source is resolved and one where it is unresolved by the 75\arcsec\ LWS beam.  The resolved subset contains 46 galaxies and the unresolved subset contains 181 galaxies.  The statistical trends in the unresolved subset are examined, and the following results are compared to earlier studies (Malhotra et al. 1997, 2001; Leech et al. 1999; Luhman et al. 1998, 2003; Negishi et al. 2001):

\noindent 1. The LWS continuum agrees with fluxes predicted from {\it IRAS} data and the spectral energy distribution models of Dale \& Helou (2002) to within 25\% at 52\m, 57\m, 63\m, 88\m, 122\m, 145\m, 158\m, and 170\m.

\noindent 2. The \CII~158\m/{\it FIR} ratio peaks for normal, star-forming galaxies with 60\m/100\m\ ratios of 0.3--0.6 and {\it FIR/B} ratios of 0.1--1.  The \CII/{\it FIR} ratio in quiescent galaxies with 60\m/100\m\ ratios less than 0.3 and {\it FIR/B} ratios less than 0.1 is consistent with normal, star-forming galaxies.  The \CII~158\m/{\it FIR} ratio decreases with increasing dust temperatures (60\m/100\m\ $> 0.6$) and infrared to blue ratio ({\it FIR/B} $> 1$).   
 
\noindent 3. The \OI~63\m/{\it FIR} ratio shows no obvious correlation with 60\m/100\m\ and a decrease as a function of {\it FIR/B}.

\noindent 4. The ratio \NII~122\m/{\it FIR} shows a similar correlation as \CII~158\m/{\it FIR}, decreasing as the 60\m/100\m\ and {\it FIR/B} ratios increase.  The \NII~122\m/\CII~158\m\ shows no correlation with either the 60\m/100\m\ or {\it FIR/B} ratio, indicating that a large fraction of \CII~158\m\ may arise from \HII\ regions.

\noindent 5. In contrast to \CII~158\m/{\it FIR}\ and \NII~122\m/{\it FIR}), the \OIII~88\m/{\it FIR} ratio increases as the 60\m/100\m\ ratio increases.  This increase might be due to the higher density of \HII\ regions found in galaxies with warmer far-infrared colors.  The \OIII~88\m/{\it FIR} ratio, however, decreases with increasing {\it FIR/B} ratio.  

\noindent 6. The \OI~63\m/\CII~158\m\ ratio increases as the 60\m/100\m\ ratio increases, but shows no correlation with {\it FIR/B}.  In warmer galaxies (60\m/100\m\ $> 0.8$), \OI~63\m\ becomes more important than \CII~158\m\ in cooling the interstellar medium.  

\noindent 7. The (\OI~63\m\ + \CII~158\m)/{\it FIR} ratio is a measure of the gas heating efficiency in PDRs, and shows only a slight decrease with increasing 60\m/100\m\ ratio for spirals but no decrease for the unresolved galaxies as a whole.  The (\OI~63\m\ + \CII~158\m)/{\it FIR} ratio does, however, decrease with increasing {\it FIR/B} ratio for the unresolved subset of galaxies as a whole.  

\noindent 8. The \OIII~88\m/\CII~158\m\ ratio increases with increasing 60\m/100\m\ ratio.  This is due to the dramatic falloff of \CII~158\m\ emission in galaxies showing warmer far-infrared emission.  The \OIII~88\m/\CII~158\m\ ratio decreases with increasing {\it FIR/B} ratio.

\noindent 9. The \OIII~88\m/\OI~63\m\ ratio has no correlation with the 60\m/100\m\ ratio.  The \OIII~88\m/\OI~63\m\ ratio decreases slightly with increasing {\it FIR/B}.


These data provide a framework through which the interstellar medium of these galaxies may be studied in the future.  

\acknowledgements 
We thank several people for their contributions: Harold Corwin for reclassifying the galaxies in this sample according to the RC3 catalog; Steve Lord, Tom Jarrett, and Alessandra Contursi for helpful discussions; Heather Maynard for her suggestions and support; Pat Patterson and Niles McElveney for assistance with the preparation of the manuscript; and an anonymous referee for many helpful suggestions.  The data for this project are based on observations with the Infrared Space Observatory, an ESA project with instruments funded by ESA member states (especially the PI countries: France, Germany, the Netherlands, and the United Kingdom) with the participation of ISAS and NASA.  The {\it ISO} Spectral Analysis Package (ISAP) is a joint development by the LWS and SWS Instrument Teams and Data Centers.  Contributing institutes are CESR, IAS, IPAC, MPE, RAL, and SRON.  This research has made use of the NASA/IPAC Extragalactic Database that is operated by the Jet Propulsion Laboratory, California Institute of Technology, under contract with the National Aeronautics and Space Administration (NASA). This work has made use of data services of the InfraRed Science Archive (IRSA) at the Infrared Processing and Analysis Center/California Institute of Technology, funded by the National Aeronautics and Space Administration (NASA).

\appendix{Appendix:  The Extended Source Correction}

The flux calibration of the LWS instrument is based on observations of Uranus, a point source in the LWS aperture.  The telescope is diffraction limited at about 110\m, beyond which a fraction of the flux of an on-axis point source may be diffracted out of the standard aperture.  Significant diffraction loss does not occur in sources that appear extended in the LWS beam.  Therefore, an extended source correction must be applied to put these fluxes on a point source calibration scale.  In order to apply this correction, the LWS beam of each detector and the telescope PSF must be well known.  The extended source correction also assumes that the source is infinitely extended and uniformly bright.  None of these galaxies are either, so applying the correction requires great caution.  From tests done by the LWS instrument team, the extended source correction works well for sources larger than 3--4\arcmin\ (Gry et al. 2003) when compared to {\it IRAS} at 100\m.   

The fluxes quoted in Tables~2, 4, and 5 are those which have been measured according to the point source flux calibration.  The current understanding of the LWS beam is still not complete, and the extended source correction may change in the future as more work is done.  Therefore, the extended source correction has not been applied to the line fluxes listed in this paper, but those galaxies that may require an extended source correction have been noted in Tables~2, 4, and 5.  The most up-to-date effective apertures and corrections are listed in Table~7, taken from the ISO LWS Handbook Volume~III (Gry et al. 2003).

                                                                                                                   

\clearpage
\begin{deluxetable}{lcc}
\def\a{\tablenotemark{a}}
\tabletypesize{\scriptsize}
\def\p{$\pm$}
\tablenum{3}
\label{tab:fir}
\tablecaption{Detected Far-Infrared Lines}
\tablewidth{0pc}
\tablehead{
\colhead{Line}&
\colhead{Wavelength}  &
\colhead{Transition}  
\\
\colhead{}      &
\colhead{(\micron)}  &
\colhead{}  
}
\startdata
[OIII]            &                 51.82 &                           $^3P_2 \rightarrow\ ^3P_1$ \cr
OH\a              &           52.93/53.06 &  $^2\pi_{3/2} J=11/2 \rightarrow\ ^2\pi_{3/2} J=9/2$ \cr
                  &           53.26/53.35 &  $^2\pi_{1/2} J= 3/2 \rightarrow\ ^2\pi_{3/2} J=3/2$ \cr
[NIII]            &                 57.32 &                    $^2P_{3/2}\rightarrow\ ^2P_{1/2}$ \cr
H$_2$O            &                 58.70 &                             o-H$_2$O $4_{32}-3_{21}$ \cr
[OI]              &                 63.18 &                           $^3P_1 \rightarrow\ ^3P_2$ \cr
OH\a              &           65.13/65.28 &  $^2\pi_{3/2} J= 9/2 \rightarrow\ ^2\pi_{3/2} J=7/2$ \cr
H$_2$O\a          &                 66.44 &                             o-H$_2$O $3_{30}-2_{21}$ \cr
\enddata                                                                                                                           
\tablenotetext{a}{\footnotesize Line detected cannot be accurately identified due to the resolution of the LWS.  The line is one or a blend of the listed transitions.}
\end{deluxetable}

\begin{deluxetable}{lccccccccccccccc}
\rotate
\def\a{\tablenotemark{a}}
\def\b{\tablenotemark{b}}
\def\c{\tablenotemark{c}}
\tabletypesize{\scriptsize}
\def\p{$\pm$}
\tablenum{5}
\label{tab:line_fluxes}
\tablecaption{Molecular and Unidentified Line Fluxes}
\tablewidth{0pc}
\tablehead{
\colhead{Galaxy}&
\colhead{TDT\a}  &
\colhead{R.A.}  &
\colhead{Dec.}  &
\colhead{OH}  &
\colhead{H$_2$O}  &
\colhead{OH} &
\colhead{H$_2$O} &
\colhead{unidentified} &
\colhead{H$_2$O}  &
\colhead{OH}  &
\colhead{OH}  &
\colhead{H$_2$O}  &
\colhead{H$_2$O}  &
\colhead{OH}  &
\colhead{OH/CO}  
\\
\colhead{}&
\colhead{}  &
\colhead{(J2000.0)}  &
\colhead{(J2000.0)}  &
\colhead{53\m}  &
\colhead{59\m}  &
\colhead{65\m} &
\colhead{67\m} &
\colhead{74\m} &
\colhead{75\m}  &
\colhead{79\m}  &
\colhead{84\m}  &
\colhead{101\m}  &
\colhead{108\m}  &
\colhead{119\m}  &
\colhead{163\m}  
}
\startdata
NGC 0253 & 24701103\b &  00 47 33.3 & $-$25 17 18 &        &        &        &        &      &        &        &         &        &        & 11.76\c  &      \cr
NGC 0253 & 56901708\b &  00 47 33.3 & $-$25 17 18 & 7.12\c &        &        &        &      &        &        &         &        &        & 11.63\c  &      \cr
NGC 0253 & 79900508\b &  00 47 33.3 & $-$25 17 18 &        &        &        &        &      &        &        &         &        &        &          & 2.24 \cr
NGC 0891 & 61100256\b &  02 22 33.0 & $+$42 20 55 &        &        &        &        &      &        &        &         &        &        & ~1.11\c  &      \cr
NGC 0891 & 61401144\b &  02 22 33.2 & $+$42 20 58 &        &        &        &        &      &        &        &         &        &        & ~1.33\c  &      \cr
NGC 1068 & 82000605   &  02 42 40.6 & $-$00 00 48 &        &        &        &        &      &        &  1.36  &         &        &        &          &      \cr
NGC 1068 & 82000704   &  02 42 40.6 & $-$00 00 48 &        &        &        &        & 1.32 &        &        &         &        &        & ~1.31    &      \cr
NGC 1068 & 82001106   &  02 42 40.6 & $-$00 00 48 &        &        &        &        & 1.24 &        &        &         &        &        & ~1.60    &      \cr
\enddata    
\tablecomments{Units of right ascension are hours, minutes, and seconds, and units of declination are degrees, arcminutes, and arcseconds.  Line fluxes are in units of 10$^{-15}$~W~m$^{-2}$.  Upper limits are also available via L01 range scans, e.g., Arp~220 and Mkn~231.}
\tablenotetext{a}{\footnotesize TDT is the ISO observation number. It has the form RRRSSSOO where RRR=revolution number, SSS=sequence number within the revolution, and OO=observation number within an observer's proposal.}
\tablenotetext{b}{\footnotesize Observation considered an extended source in this paper. May require an extended source correction.}
\tablenotetext{c}{\footnotesize Flux measured is in absorption rather than emission.}                                 
\end{deluxetable}

\begin{deluxetable}{lcrrrrrrr}
\def\a{\tablenotemark{a}}
\def\b{\tablenotemark{b}}
\def\c{\tablenotemark{c}}
\def\d{\tablenotemark{d}}
\def\e{\tablenotemark{e}}
\def\f{\tablenotemark{f}}
\def\g{\tablenotemark{g}}
\tabletypesize{\scriptsize}
\def\p{$\pm$}
\tablenum{6}
\label{tab:serendipitous}
\tablecaption{Serendipitous Galactic \CII~158\m\ Detections}
\tablewidth{0pc}
\tablehead{
\colhead{Galaxy}&
\colhead{TDT\a}  &
\colhead{$l$}  &
\colhead{$b$}  &
\colhead{c$z$}  &
\colhead{Galactic}  &
\colhead{Galactic} &
\colhead{extragalactic} &
\colhead{contamination\f} 
\\
\colhead{}&
\colhead{}  &
\colhead{(\degr)}  &
\colhead{(\degr)}  &
\colhead{(km~s$^{-1}$)}  &
\colhead{prediction\b}  &
\colhead{observation\c} &
\colhead{observation\d} &
\colhead{(\%)} 
}
\startdata
IC 10        &45700606   &118.89&  $-$3.23& $-$344&  25.38&  20.73\p0.48&  70.93\g& 23.4 \cr
             &45700607   &119.06&  $-$3.45&       &  23.76&  12.44\p0.48&         &      \cr
Maffei 2     &85800683\e &136.48&  $-$0.55&  $-$17& 148.97&  27.47\p0.87& 112.62  & 24.4 \cr
NGC 1156     &65300185   &156.37& $-$29.28&    375&   2.92&   2.51\p0.58&  17.54  & 14.3 \cr
UGC 02855    &62902697   &136.91& $+$12.33&   1202&   6.70&   4.85\p0.58&  50.51  &  9.6 \cr
NGC 1569     &64600490   &143.61& $+$11.31& $-$104&   7.29&   8.22\p0.86&  65.99  & 12.5 \cr
3C 120       &80901988   &190.37& $-$27.40&   9896&   3.11&   3.95\p0.29&   3.66  &    0 \cr
NGC 1614     &85501010   &204.45& $-$34.38&   4778&   2.53&   4.44\p0.39&  21.80  &    0 \cr
\enddata   
\tablecomments{[CII]~158m\ intensities are in $10^{-6}~{\rm ergs}~{\rm cm}^{-2}~{\rm s}^{-1}~{\rm sr}^{-1}$ assuming a 75\arcsec\ aperture diameter.}
\tablenotetext{a}{\footnotesize TDT is the ISO observation number. It has the form RRRSSSOO where RRR=revolution number, SSS=sequence number within the revolution, and OO=observation number within an observer's proposal.}
\tablenotetext{b}{\footnotesize Milky Way [CII]~158\m\ predicted by Equation~\ref{eq:CII}.}
\tablenotetext{c}{\footnotesize Milky Way [CII]~158\m\ measured by ISOLWS.}
\tablenotetext{d}{\footnotesize [CII]~158\m\ measured for external target galaxy.}
\tablenotetext{e}{\footnotesize Observation considered an extended source in this paper. May require an extended source correction.}
\tablenotetext{f}{\footnotesize Percent of observed [CII] arising from the Milky Way.  This foreground contamination has been removed from the [CII] fluxes listed in Table~4 for these galaxies.}
\tablenotetext{g}{\footnotesize An average value measured in the LWS aperture from multiple observations.}
\end{deluxetable}

\begin{deluxetable}{ccc}
\tabletypesize{\scriptsize}
\def\p{$\pm$}
\tablenum{7}
\label{tab:serendipitous}
\tablecaption{Extended Source Aperture Corrections}
\tablewidth{0pc}
\tablehead{
\colhead{Detector}&
\colhead{Aperture Radius}  &
\colhead{Correction} 
\\
\colhead{}&
\colhead{(\arcsec)}  &
\colhead{} 
}
\startdata
SW1 & 39& 0.88 \cr
SW2 & 42& 0.88 \cr
SW3 & 43& 0.84 \cr
SW4 & 41& 0.73 \cr
SW5 & 40& 0.70 \cr
LW1 & 38& 0.69 \cr
LW2 & 39& 0.69 \cr
LW3 & 35& 0.62 \cr
LW4 & 34& 0.55 \cr
LW5 & 33& 0.48 \cr
\enddata   
\tablecomments{The fluxes listed in Tables 3 and 4 for extended sources are based on a point source calibration.  The multiplicative correction listed here may need to be applied to these fluxes.}
\end{deluxetable}

 

\clearpage
\begin{figure}
 \plotone{f1.eps}
 \caption{Aitoff projection of the galaxies selected for this sample.  The galaxies in this sample are distributed across the sky.  Galaxies unresolved by the LWS in the far-infrared are displayed with crosses.  Resolved galaxies by the LWS in the far-infrared are shown with open squares.  The clump of galaxies at RA, Dec (70\degr, 280\degr) are members of the Virgo Cluster.}
\label{fig:aitoff}
\end{figure}

\begin{figure}
 \caption{Three LWS spectra with both Milky Way \CII~158\m\ and C$^+$ emission at the redshift of the observed galaxy is plotted along with the corresponding 1\degr$\times$1\degr\ {\it IRAS} 100\m\ images.  The LWS observation for each galaxy was located at the center of each {\it IRAS} image.  The line through the spectra is a best fit using a linear baseline and two instrumental width Gaussians.}
\label{fig:CII_MW}
\end{figure}

\clearpage

\clearpage

\clearpage
\begin {thebibliography}{dum}
\bibitem[]{}Abel, N.P., Ferland, G.,J., Shaw, G., \& van Hoof, P.A.M. 2005, \apjs, 161, 65
\bibitem[]{}Bennett, C.L. et al. 1994, \apj, 434, 587
\bibitem[]{}Boreiko, R.T. \& Betz, A.L. 1997, \apjs, 111, 409
\bibitem[]{}Boulanger, F., Beichman, C., D\'esert, F.X., Helou, G., Perault, M., \& Ryter, C. 1988, \apj, 332, 328 
\bibitem[]{}Bradford, C.M. et al. 1999, in ``The Universe as Seen by ISO'', ed. Cox, P. \& Kessler, M.F., ESA SP-427 (Noordwijk, Netherlands), 861
\bibitem[]{}Carral, P., Hollenbach, D.J., Lord, S.D., Colgan, S.W.J., Haas, M.R., Rubin, R.H., \& Erickson, E.F. 1994, \apj, 423, 223
\bibitem[]{}Colbert J.W. et al. 1999, \apj, 511, 721
\bibitem[]{}Clegg, P. et al. 1996, \aap, 315, L38
\bibitem[]{}Colgan, S.W.J., Haas, M.R., Erickson, E.F., Rubin, R.H., Simpson, J.P., \& Russell, R.W. 1993, \apj, 413, 237
\bibitem[]{}Contursi, A. et al. 2002, \apj, 124, 751
\bibitem[]{}Crawford, M.K., Genzel, R., Townes, C.H., \& Watson, D.M. 1985, \apj, 291, 755
\bibitem[]{}Dale, D.A. et al. 2000, \apj, 120, 583
\bibitem[]{}Dale, D.A., Helou, G., Contursi, A., Silbermann, N.A., \& Kolhatkar, S. 2001, \apj, 549, 215 
\bibitem[]{}Dale, D.A. \& Helou, G. 2002, \apj, 576, 159 
\bibitem[]{}Dale, D.A., Helou, G., Brauher, J.R., Cutri, R.M., Malhotra, S., \& Beichman, C.A. 2004, 604, 565
\bibitem[]{}Dopita, M.A., Fischera, J., Sutherland, R.S., Kewley, L.J., Leitherer, C., Tuffs, R.J., Popescu, C.C., van Breugel, W., \& Groves, B.A. 2006, \apjs, 167, 177
\bibitem[]{}Duffy, P.B., Erickson, E.F., Haas, M.R., \& Houck, J.R. 1987, \apj, 315, 68
\bibitem[]{}Fischer, J. et al. 1996, \aap, 315, L97
\bibitem[]{}Fischer, J. et al. 1999, Astrophysics \& Space Science, 266, 91
\bibitem[]{}Fullmer, L. \& Londsdale, C.J. 1989, Cataloged Galaxies and Quasars Observed in the {\it IRAS} Survey, Version 2 (Pasadena:JPL)
\bibitem[]{}Goicoechea, J.R., Mart\'in-Pintado, J., \& Cernicharo, J. 2005, \apj, 619, 291
\bibitem[]{}Gonz\'alez-Alfonso, E., Smith, H.A., Fischer, J., \& Cernicharo, J. 2004, \apj, 613, 247
\bibitem[]{}Gonz\'alez-Alfonso, E., Smith, H.A., Ashby, M.L.N., Fischer, J., Spinoglio, L., \& Grundy, T.W. 2008, \apj, 675, 303
\bibitem[]{}Groves, B., Dopita, M., Sutherland, R., Kewley, L., Fischera, J., Leitherer, C., Brandl, B., \& van Breugal, W. 2008, \\apjs, in press
\bibitem[]{}Gry, C. et al. 2003, The ISO Handbook, Volume III -- LWS -- The Long Wavelength Spectrometer, Version 2.1 (European Space Agency)
\bibitem[]{}Hartmann, D. \& Burton, W.B. 1997, Atlas of Galactic Neutral Hydrogen (Cambridge: Cambridge University Press)
\bibitem[]{}Heiles, C. 1994, \apj, 436, 720
\bibitem[]{}Helou, G. 1986, \apj, 311, L33
\bibitem[]{}Helou, G., Khan, I.R., Malek, L., \& Boehmer, L. 1988, \apjs, 68, 151
\bibitem[]{}Hollenbach, D.J., Takahashi, T., \& Tielens, A.G.G.M. 1991, \apj, 377, 192
\bibitem[]{}Hollenbach, D.J. \& Tielens, A.G.G.M. 1997, \araa, 35, 179
\bibitem[]{}Hunter, D.A. et al. 2001, \apj, 553, 121
\bibitem[]{}Hutchinson, M.G., Chan, S.J., \& Sidher, S.D. 2001, in ``The Calibrations Legacy of the {\it ISO} Mission'', ed. Metcalfe, L. \& Kessler, M.F., ESA SP-481
\bibitem[]{}Kaufman, M.J., Wolfire, M.G., Hollenbach, D.J., \& Luhman, M.L. 1999, \apj, 527, 795
Kegel, W.H., Hertenstein, T., \& Quirrenbach, A. 1999, \aap, 351, 472
\bibitem[]{}Kessler, M.F. et al. 1996, \aap, 315, L27
\bibitem[]{}Kessler, M.F., Mueller, T.G., Leech, K., Arviset, C., Garcia-Lario, P., Metcalfe, L., Pollock, A., Prusti, T., \& Salama, A. 2003, The ISO Handbook, Volume I -- Mission \& Satellite Overview, Version 2.0 (European Space Agency)
\bibitem[]{}Le Petit, F., Nehm\'e, C., Le Bourlot, J., \& Roueff, E. 2006, \\apjs, 164, 506
\bibitem[]{}Leech, K.J., Völk, Heinrichsen, I., Hippelein, H., Metcalfe, L., Pierini, D., Popescu, C.C., Tuffs, R.J., \& Xu, C. 1999, \mnras, 310, 317
\bibitem[]{}Liu, X.-W. et al. 1996, \aap, 315, L257
\bibitem[]{}Lord, S.D. Hollenbach, D.J., Haas, M.R., Rubin, R.H., Colgan, S.W.J., \& Erickson, E.F. 1996, \apj, 465, 703
\bibitem[]{}Luhman, M.L., Satyapal. S., Fischer, J., Wolfire, M.G., Cox, P., Lord, S.D., Smith, H.A., Stacey, G.J., \& Unger, S.J. 1998, \apj, 504, L11
\bibitem[]{}Luhman, M.L., Satyapal. S., Fischer, J., Wolfire, M.G., Sturm, E., Dudley, C.C., Lutz, D. \& Genzel, R. 2003, \apj, 594, 758
\bibitem[]{}Madden, S.C., Geis, N., Genzel, R., Herrmann, F., Jackson, J., Poglitsch, A., Stacey, G.J., \& Townes, C.H. 1993, \apj, 407, 579
\bibitem[]{}Madden, S.C., Poglitsch, A., Geis, N., Stacey, G.J., \& Townes, C.H. 1997, \apj, 483, 200
\bibitem[]{}Malhotra, S. et al. 1997, \apj, 491, L27
\bibitem[]{}Malhotra, S. et al. 2000, \apj, 543, 634
\bibitem[]{}Malhotra, S. et al. 2001, \apj, 561, 766
\bibitem[]{}Maloney, P.R., Hollenbach, D.J., \& Tielens, A.G.G.M. 1996, \apj, 466, 561
\bibitem[]{}Meijerink, R., Spaans, M., \& Israel, F.P. 2007, \aap, 461, 793
\bibitem[]{}Nakagawa, T., Doi, Y., Yui, Y.Y., Okuda, H., Mochizuki, K., Shibai, H., Nishimura, T., \& Low, F.J. 1995, \apj, 455, L35 
\bibitem[]{}Negishi, T., Onaka, T., Chan, K.-W., \& Roellig, T.L., 2001, \aap, 375, 566
\bibitem[]{}Oliva, E., Moorwood, A.F.M., Drapatz, S., Lutz, D., \& Sturm, E. 1999, \aap, 343, 943
\bibitem[]{}Petuchowski, S.J. \& Bennett, C.L. 1993, \apj, 405, 591
\bibitem[]{}Pierini, D., Leech, K.J., Tuffs, R.J., \& Völk, H.J. 1999, \mnras, 303, L29
\bibitem[]{}Reach, W.T., Wall, W.F., \& Odegard, N. 1998, \apj, 507, 507
\bibitem[]{}Rubin, R.H. 1985, \apjs, 57, 349
\bibitem[]{}Rubin, R.H., Simpson, J.P., Lord, S.D., Colgan, S.W.J., Erickson, E.F., \& Haas, M.R. 1994, \apj, 420, 772
\bibitem[]{}Sauty, S., Gerin, M., \& Casoli, F. 1998, \aap, 339, 19
\bibitem[]{}Sauvage, M., Tuff, R.J., \& Popescu, C.C. 2005, Space Science Reviews, Vol. 119, Issue 1-4, 313
\bibitem[]{}Shibai, H., Okuda, H., Nakagawa, T., Matsuhara, H., Maihara, T., Mizutani, K., Kobayashi, Y., \& Hiromoto, N. 1991, \apj, 374, 522
\bibitem[]{}Skinner, C.J., Smith, H.A., Sturm, E., Barlow, M.J., Cohen, R.J., \& Stacey, G.J. 1997, Nature, 386, 472
\bibitem[]{}Spinoglio, L. \& Malkan, M.A. 1992, \apj, 399, 504
\bibitem[]{}Spinoglio, L., Malkan, M.A., Smith, H.A., Gonz\a'lex-Alfonso, E., \& Fischer, J. 2005, \apj, 623, 123
\bibitem[]{}Sturm, E. et al. 1998, Astronomical Data Analysis Software and Systems VII, A.S.P. Conference Series, Vol. 145, eds. R. Albrecht, R.N. Hook, \& H.A. Bushouse, p. 161
\bibitem[]{}Stacey, G.J., Geis, N., Genzel, R., Lugten, J.B., Poglitsch, A., Sternberg, A., \& Townes, C.H. 1991, \apj, 373, 423
\bibitem[]{}Stickel, M., Lemke, D., Klaas, U., Beichman, C.A., Rowan-Robinson, M., Efstathiou, A., Bogun, S., Kessler, M.F., \& Richter, G. 2000, \aap, 359, 865
\bibitem[]{}Swinyard, B.M., Burgdorf, M.J., Clegg, P.E., Davis, G.R., Griffin, M.J., Gry, C., Leeks, S.J., Lim, T.L., Pezzuto, S., \& Tommasi, E. 1998, SPIE, 3354, 888
\bibitem[]{}Tielens, A.G.G.M. \& Hollenbach, D.J. 1985, \apj, 291, 722
\bibitem[]{}Unger, S.J. et al. 2000, \aap, 355, 885
\bibitem[]{}Varberg, T.D. \& Evenson, K.M. 1992, \apj, 385, 763
\bibitem[]{}Verma, A., Charmandaris, V., Klaas, U., Lutz, D., \& Haas, M. 2005, Space Science Reviews, Vol. 119, Issue 1-4, 355
\bibitem[]{}Wolfire, M.G., Tielens, A.G.G.M., \& Hollenbach, D.J. 1990, \apj, 358, 116
\bibitem[]{}Wright, E.L. et al. 1991, \apj, 381, 200 
\end {thebibliography}
\end{document}